\documentclass[fleqn,usenatbib]{mnras}

\usepackage{newtxtext,newtxmath}
\usepackage[T1]{fontenc}
\usepackage{ae,aecompl}

\usepackage{graphicx}	
\usepackage{amsmath}
\usepackage{afterpage}
\usepackage{longtable}
\usepackage{siunitx}
\usepackage{microtype}
\usepackage{url}

\title[V392\,Persei:\ a ${\gamma}$-ray bright nova]{V392\,Persei:\ a $\boldsymbol{\gamma}$-ray bright nova eruption from a known dwarf nova}

\author[F. J. Murphy-Glaysher et al.]{
F. J. Murphy-Glaysher,$^{1}$\thanks{E-mail: F.J.MurphyGlaysher@2018.ljmu.ac.uk}
M. J. Darnley,$^{1}$
{\'E}. J. Harvey,$^{1}$
A. M. Newsam,$^{1}$
K. L. Page,$^{2}$
\newauthor
S. Starrfield,$^{3}$
R. M. Wagner,$^{4}$
C. E.  Woodward,$^{5}$
D. M. Terndrup,$^{4}$
S. Kafka,$^{6,7}$
\newauthor
T. Arranz Heras,$^{6}$ 
P. Berardi,$^{8}$ 
E. Bertrand,$^{8}$
R. Biernikowicz,$^{6}$
C. Boussin,$^{8,9}$
D. Boyd,$^{6}$
\newauthor
Y. Buchet,$^{8}$
M. Bundas,$^{6}$
D. Coulter,$^{6}$
D. Dejean,$^{8}$
A. Diepvens,$^{6}$
S. Dvorak,$^{6}$
J. Edlin,$^{8}$
\newauthor
T. Eenmae,$^{6,10}$
H. Eggenstein,$^{6}$
R. Fournier,$^{6}$
O. Garde,$^{8,11}$
J. Gout,$^{6}$
D. Janzen,$^{6,12}$
P. Jordanov,$^{6}$
\newauthor
H. Kiiskinen,$^{6}$
D. Lane,$^{6}$
R. Larochelle,$^{6,12}$
R. Leadbeater,$^{8}$
D. Mankel,$^{6}$
G. Martineau,$^{8}$
\newauthor
I. Miller,$^{6}$
R. Modic,$^{6}$
J. Montier$^{8}$
M. Morales Aimar,$^{6}$
E. Muyllaert,$^{6}$
R. Naves Nogues,$^{6}$
\newauthor
D. O'Keeffe,$^{6}$
A. Oksanen,$^{6}$
M. Pyatnytskyy,$^{6}$
R. Rast,$^{6,12}$
B. Rodgers,$^{6,12}$
D. Rodriguez Perez,$^{6}$
\newauthor
F. Schorr,$^{6}$
E. Schwendeman,$^{6}$
S. Shadick,$^{6,12}$
S. Sharpe,$^{6}$ 
F. Sold\'an Alfaro,$^{6,13}$ 
T. Sove,$^{6,12}$ 
\newauthor
G. Stone,$^{6}$
T. Tordai,$^{6}$
R. Venne,$^{6}$
W. Vollmann,$^{6}$
M. Vrastak,$^{6}$
K. Wenzel$^{6}$
\\
$^{1}$Astrophysics Research Institute, Liverpool John Moores University, IC2, Liverpool Science Park, Brownlow Hill, Liverpool L3 5RF, UK\\
$^{2}$School of Physics \& Astronomy, University of Leicester, LEI 7RH, UK\\
$^{3}$School of Earth and Space Exploration, Arizona State University, Box 871404, Tempe, AZ 85287-1404, USA\\
$^{4}$Department of Astronomy, The Ohio State University, Columbus, OH 43210, USA\\
$^{5}$Minnesota Institute for Astrophysics, School of Physics \& Astronomy, 116 Church Street SE, University of Minnesota, Minneapolis, MN 55455, USA\\
$^{6}$American Association of Variable Star Observers, 49 Bay State Rd, Cambridge, MA 02138, USA\\
$^{7}$American Meteorological Society, 45 Beacon Street, Boston, MA 02108-3693, USA\\
$^{8}$Astronomical Ring for Access to Spectroscopy --- \url{http://www.astrosurf.com/aras}\\
$^{9}$Observatoire de l'Eridan et de la Chevelure de B\'{e}r\'{e}nice, France\\
$^{10}$Tartu Observatory, University of Tartu, Estonia\\
$^{11}$Observatoire de la Tourbi\`{e}re, France\\
$^{12}$Department of Physics and Engineering Physics, University of Saskatchewan, Saskatoon, SK, S7N 5E2, Canada\\
$^{13}$Seville University, Seville, Spain
}

\date{Accepted 2022 June 6. Received 2022 June 1; in original form 2022 April 20}

\pubyear{2022}

\begin{document}
\label{firstpage}
\pagerange{\pageref{firstpage}--\pageref{lastpage}}
\maketitle

\begin{abstract} 
V392 Persei is a known dwarf nova (DN) that underwent a classical nova eruption in 2018. Here we report ground-based optical, \textit{Swift} UV and X-ray, and \textit{Fermi}-LAT $\gamma$-ray observations following the eruption for almost three years. V392 Per is one of the fastest evolving novae yet observed, with a $t_2$ decline time of 2 days. Early spectra present evidence for multiple and interacting mass ejections, with the associated shocks driving both the $\gamma$-ray and early optical luminosity. V392 Per entered Sun-constraint within days of eruption. Upon exit, the nova had evolved to the nebular phase, and we saw the tail of the super-soft X-ray phase. Subsequent optical emission captured the fading ejecta alongside a persistent narrow line emission spectrum from the accretion disk. Ongoing hard X-ray emission is characteristic of a standing accretion shock in an intermediate polar. Analysis of the optical data reveals an orbital period of $3.230\pm0.003$ days, but we see no evidence for a white dwarf (WD) spin period. The optical and X-ray data suggest a high mass WD, the pre-nova spectral energy distribution (SED) indicates an evolved donor, and the post-nova SED points to a high mass accretion rate. Following eruption, the system has remained in a nova-like high mass transfer state, rather than returning to the pre-nova DN low mass transfer configuration. We suggest that this high state is driven by irradiation of the donor by the nova eruption. In many ways, V392 Per shows similarity to the well-studied nova and DN GK Persei.\end{abstract}


\section{Introduction}

Classical novae (CNe) are among the most luminous stellar transients, exceeded only by supernovae and gamma-ray bursts. CNe are binary systems in which a white dwarf (WD; typically composed of carbon and oxygen, or of oxygen and neon) accretes hydrogen-rich material from a donor star via an accretion disk \citep{Warner1995}. Accretion proceeds via Roche-lobe overflow for the majority of CNe; those with main sequence or sub-giant donors. For CNe with giant donors, material is accreted from the giant's wind. The accreted envelope builds in temperature and pressure until a thermonuclear runaway occurs \citep{Starrfield1976,Starrfield2016,Starrfield1989}, blasting material from the WD's surface, leaving the WD and donor relatively unscathed. The CN is observed as a rapid increase in optical luminosity of 10--15 magnitudes, followed by a slower decline.

CNe are a sub-type of cataclysmic variable (CV); a class that includes dwarf novae (DNe) and nova-likes (NL). DN outbursts are less luminous than CN eruptions and are powered by the release of gravitational potential energy, which can occur when hydrogen-rich material in the accretion disk is suddenly deposited onto the WD. DN outbursts are produced
in systems where the accretion rate ($\dot{M}$) is lower than the critical rate \citep[see their Equation~2]{Smak1983}, due to thermal or tidal instabilities within the disk \citep{Osaki1996}. For a given  disk radius, CVs with high $\dot{M}$ produce hot, stable disks -- the NL systems, that do not show DN outbursts. 

\citet{Abdo2010} first reported detection of $\gamma$-ray emission from a nova; the V407\,Cygni ejecta shocked surrounding circumstellar wind, accelerating leptons to relativistic velocities and emitting $\gamma$-ray photons of energy $>100$\,MeV. Since that initial discovery, $\gamma$-ray signatures have been exhibited in increasing numbers of classical novae \citep*[see][for recent reviews]{Aydi2020ApJ,2021ARA&A..59..391C}. Several $\gamma$-ray detected novae occurred in systems with red giant donors: V407\,Cyg, V1535\,Sco \citep{Franckowiak2018}, and the recurrent novae V745\,Sco \citep{Cheung2014,Bannerjee2014}, V3890\,Sgr \citep*{Buson2019} and RS\,Oph \citep*{Cheung2021}. In these systems, the shocks generating the $\gamma$-rays are likely to originate in collisions between the nova ejecta and the dense red-giant winds and circumbinary material. However, the other $\gamma$-ray emitting novae have main sequence companions and are unlikely to be surrounded by dense winds. In these systems, the shocks are proposed to be due to interaction between multiple ejection components \citep{Aydi2020ApJ}. 
Studies suggest that the $\gamma$-ray  and optical emission can show correlated peaks, with the shocks driving the optical emission \citep{Ackermann14}. \citet{Aydi2020} demonstrated that shock-powered emission was responsible for the bulk of the luminosity of V906\,Car, with multiple simultaneous $\gamma$-ray and optical flares.

A number of CVs have been observed to undergo both CN eruptions and DN outbursts. GK\,Per \citep{1986A&A...160..367B,Zemko2017}, V446\,Her \citep*{Honeycutt2011}, RR\,Pic, V1047\,Cen, and V606 Aql \citep{Kato2021} are CNe that subsequently underwent DN outbursts. Z\,Cam, AT\,Cnc, and 2MASS\,J17012815-4306123 (Nova Sco 1437) are known DNe surrounded by proposed ancient CN shells \citep{Shara2007,Shara2012,Shara2017N}. V1213\,Cen and V1017\,Sgr exhibited DN outbursts six and eighteen years, respectively, before a CN; V1017\,Sgr also showed post-nova DN outbursts \citep{Mroz2016}. The nebula Te\,11, with a DN at its centre, was proposed to be the shell of an ancient nova eruption, rather than a planetary nebula \citep{Miszalski2016}. 

V392\,Persei was a known CV with a few observed DN outbursts, with quiescent magnitudes of $15.0<m_\mathrm{pg}<17.0$ \citep{Downes1993} and $V >17$ \citep{Zwitter1994}. Its CN eruption was discovered on 2018 Apr 29 (UT) by Y.\ Nakamura, with an unfiltered brightness of 6.2 mag \citep{Wagner2018}. The following day, $\gamma$-ray emission was detected from V392\,Per \citep*[$>6\sigma$;][]{Li2018}, with detections continuing for 11 days \citep{Gordon2021,2022arXiv220110644A}. Non-thermal synchrotron emission during early radio observations \citep{2021ApJS..257...49C} provided further support for the presence of shocks during the eruption. The system is proposed to host an evolved donor similar to the sub-giant donors of U\,Sco and GK\,Per, or the low-luminosity giant donor of M31N\,2008-12a \citep{DarStar18}. Potential orbital periods of 3.4118\,days \citep*{Munari2020} and 3.21997\,days \citep{Schaefer2021} are consistent with an evolved donor.

In this paper, we present panchromatic data from the 2018 nova eruption of V392\,Per and its subsequent evolution. In Section \ref{sec:Data} we describe our observational dataset. We present the photometry and spectra in Section \ref{sec:Results} and the spectral analysis in Section \ref{sec:SpecAnalysis}. We discuss our results in Section \ref{sec:Disc} and summarise our findings in Section \ref{sec:Conc}.

\section{Data}\label{sec:Data}

Observations of V392\,Per were obtained by the fully robotic 2.0\,m Liverpool Telescope \citep[LT;][]{Steele2004} on La Palma. LT images were taken with the IO:O instrument \citep{Smith2017} through $u'BVr'i'z'$ filters. Additional $i'$-band photometry was collected using both Las Cumbres Observatory \citep[LCOGT;][]{2013PASP..125.1031B} 1.0\,m Telescopes at McDonald Observatory in Fort Davis, Texas. We also used optical photometry from the American Association of Variable Star Observers \citep[AAVSO;][]{AAVSO}. 

The LT and LCOGT data were reduced using standard tools within PyRAF \citep{Pyraf} and aperture photometry was performed using {\tt qphot}. The data were calibrated against 25 reference stars in the field (see Table~\ref{tab:refphot}), selected from the Pan-STARRS catalogue \citep[DR1;][]{Chambers2016}. The reference stars had $g'r'i'z'$ magnitudes $14<m<22$, and were sufficiently distant from  other stars. Pan-STARRS photometry was converted to Johnson $BV$ and Sloan-like $r'i'z'$ using relations in \cite{Tonry2012}. A single star in a \textit{Swift} observation of the field was utilised to calibrate the $u'$-band photometry. For comparison with the LT/LCOGT data, the AAVSO photometry was converted to the AB system using relations from \citet{Blanton2007}. Due to the typically larger or unknown uncertainties on the AAVSO data and the large number of independent observers, we opted not to apply colour corrections to these data.

LT spectra were collected using SPRAT \citep[with a spectral resolution $R\approx350$;][blue-optimised mode]{Piascik2014} and FRODOSpec \citep*{Barnsley2012}. Groups of three or five exposures were taken at each epoch. Cosmic rays were removed by a two-stage process involving interactive interpolation and exposure combination with the IRAF routine {\tt scombine} \citep{IRAF}. The resolving power of the FRODOSpec red arm was $R\approx2200$ or $R\approx5300$, and the blue arm was $R\approx2600$ or $R\approx5500$. These spectra were reduced using the LT pipeline; producing bias subtracted, flat-fielded, wavelength calibrated, sky-subtracted products. 

Relative flux calibration of the SPRAT spectra was conducted with 78 observations of the spectrophotometric standard Hiltner\,102 \citep{Stone1977}. Relative flux calibration of the FRODOSpec spectra was performed using observations of the spectrophotometric standard stars Hiltner\,102 and BD+33\,2642 \citep{1990AJ.....99.1621O} for the higher and lower resolution modes, respectively.

Optical spectra of V392 Per were obtained with the 2.4\,m Hiltner telescope of the Michigan-Dartmouth-MIT (MDM) Observatory on Kitt Peak. The Hiltner spectra were obtained using the Ohio State Multi-Object Spectrograph \citep[OSMOS;][]{Martini2011}.  A 1\farcs2 wide entrance long-slit combined with high-efficiency, low-resolution  blue and red optimized VPH grisms and a 4096 $\times$ 4096 pixel STA CCD, was employed covering either the 3980–6860 \AA\ spectral region (the inner slit position) or the 3200–9000 \AA\ region (the centre slit position).  In both cases, the nominal spectral resolution $\Delta\lambda$ was about 3.8 \AA\ corresponding to $R \simeq 1600$ at 6000 \AA.  Spectra of HgArNeXe spectral line lamps and of a quartz-halogen lamp were obtained to provide wavelength calibration and to flat field the detector respectively.  In general, several spectra were obtained of the standard stars Hiltner 102 or G191-B2B \citep{Stone1977} on each night to remove the instrumental response function and provide relative flux calibration. The spectra were reduced using IRAF packages to subtract the bias overscans from each of the four quadrants of the detector and produce flat-fielded, one dimensional wavelength and flux calibrated extracted spectra.

We obtained optical spectra of V392 Per at two epochs with the 8.4 m Large Binocular Telescope \citep[LBT;][]{2008SPIE.7012E..03H} and the Multi Object Double Spectrograph \citep[MODS;][]{2010SPIE.7735E..0AP}. MODS utilizes 
separate and optimized blue and red channel spectrographs. Observations
at both epochs consisted of simultaneous blue and red spectra with R $\sim1850$ and R $\sim 2300$ respectively, covering the 3249–10100 \AA\ spectral region. 
Spectra of HgArNeXe lamps and an internal quartz–halogen lamp were used to determine the wavelength calibration and to flat–field the spectra. Standard stars, including G191-B2B, were used to measure the instrumental response 
function and to provide relative flux calibration. The spectra were reduced 
using custom routines to bias-subtract and flat–field the data and then 
using IRAF \citep{IRAF} for spectral extraction, wavelength and flux calibration. The wavelength scale of the 2019 data has been shifted by 2 \AA\ post-reduction to ensure that the position of the H$\alpha$ emission lines coincide.

All aforementioned spectra were absolute flux calibrated using interpolated $BVr'i'$ photometry (see Section~\ref{sec:LCfitting}).  Spectra were corrected for heliocentric velocity and dereddened using $E\left(B-V\right)=0.7$ (see Section \ref{sec:reddening}).  Additional spectra were retrieved from the Astronomical Ring for Access to Spectroscopy database\footnote{\url{http://www.astrosurf.com/aras/Aras_DataBase/DataBase.htm}} \citep[ARAS;][]{ARAS2019}. These spectra were only used to measure the P\,Cygni velocities from the Balmer lines. All spectra are listed in Table~\ref{tab:Specobs}.

Neil Gehrels {\it Swift} Observatory \citep{Gehrels2004} observations of V392\,Per (target IDs:\ 10734 and 10773) were obtained using the X-ray Telescope \cite[XRT;][]{Burrows2005} and UV/Optical Telescope \citep[UVOT;][]{Roming2005}. An initial observation on 2018 July 20, upon emerging from Sun constraint, was taken to ascertain the suitability of the target for the UVOT UV-grism, and determined that it was too faint. Subsequent observations were taken approximately weekly with XRT, initially in automatic mode, before switching to XRT in photon counting (PC) mode. In October 2018, observations switched to every two weeks until April 2019, from July 2019 observations moved to a four-weekly cadence, and monthly from 2020 January--April. The final observation was taken in 2020 August. {\it Swift} data were processed and analysed using the standard HEASoft tools and relevant calibration files. 

XRT analysis utilised the full event range of grades 0--12. At no point did the data suffer from pile-up. A circular extraction region of 10--15 pixels ($2.36^{\prime\prime}\,\mathrm{pixel}^{-1}$) was used, depending on the source brightness; the background was estimated from a 60 pixel radius circle, offset from, but close to, the source. Upon examination of the hardness ratio (HR), it was clear that there was no rapid spectral evolution. The on-line XRT product generator\footnote{\url{https://www.swift.ac.uk/user_objects}} \citep{Evans2009} was used to extract spectra over a number of intervals, during each of which the HR remained approximately constant.

UVOT analysis utilised the updated sensitivity calibrations released in 2020. Magnitudes were estimated using {\tt uvotsource}, with a $3^{\prime\prime}$ radius source extraction region to avoid possible low-level contamination from a nearby source, and $20^{\prime\prime}$ background circle.

\begin{figure*}
\begin{center}
	\includegraphics[width=0.8\textwidth]{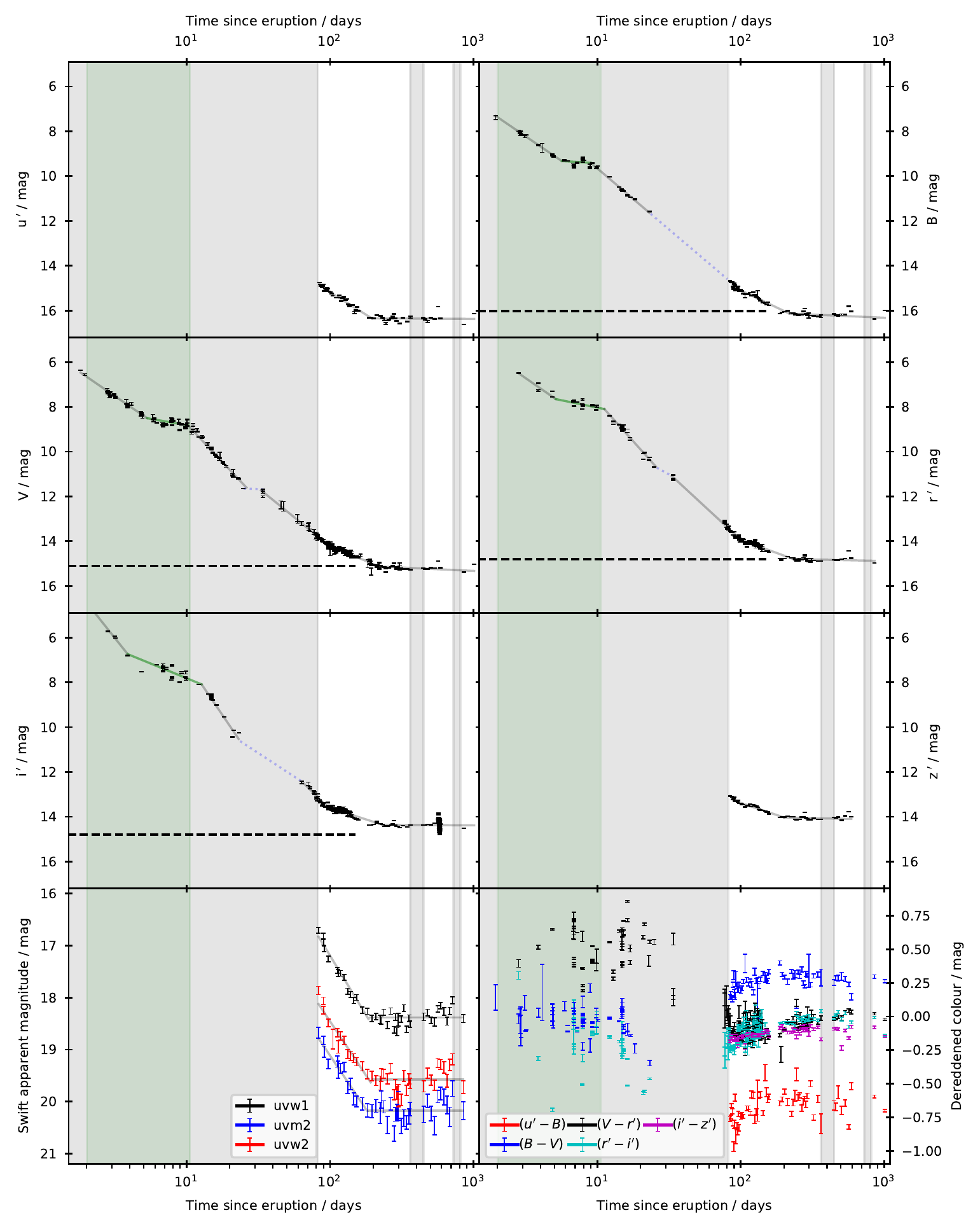}
	\end{center}
    \caption{Light curves of the nova eruption of V392\,Per from LT, LCOGT, AAVSO, and {\it Swift} observations.  Grey regions indicate the {\it Swift} Sun constraints, and the green region demarks the epoch of $\gamma$-ray detection by {\it Fermi}-LAT. Broken power laws have been fitted to each light curve (see Section~\ref{sec:LCfitting}). The horizontal dashed line indicates photometry of a nearby ($9^{\prime\prime}$ away) star in the field. {\bf Bottom-right}: Dereddened colour evolution using $E(B-V)=0.7$\,mag.}
    \label{fig:LCAll_log}
\end{figure*}

All photometry is recorded in Table~\ref{tab:Bphot}, and near-UV and optical light curves of  V392\,Per are presented in Figure~\ref{fig:LCAll_log}.

\section{Results}\label{sec:Results}

\subsection{Sun constraints}

Observation of V392\,Per was heavily impacted by on-sky proximity to the Sun at the time of its eruption. Early ground-based observations focussed on spectroscopic data until the system entered its Sun constraint. \textit{Swift} was constrained more severely than ground-based telescopes. The \textit{Swift} Sun constraints covered the periods from discovery up to 2018 Jul 18; 2019 Apr 25 to  Jul 20; and 2020 Apr 24 to Jul 19. The {\it Swift} constraints are shown by the vertical shaded regions in Figure \ref{fig:LCAll_log}.

\subsection{Time of eruption}\label{sec:Tnought}

The eruption of V392\,Per was discovered on 2018 April 29 by Yuji Nakamura in Kameyama \citep{Wagner2018}; who also collected the last pre-eruption observation on April 21.4627\footnote{\url{http://www.cbat.eps.harvard.edu/unconf/followups/J04432130+4721280.html}}, 8.01\,days earlier. The discovery observation is the brightest of the nova and one could assume that this coincides with onset of the eruption ($T_0$), or with maximum light ($T_\mathrm{max}$). Thus, we used the $V$-band light curve (see Figure~\ref{fig:LCAll_log}) to estimate $t_2$ and $t_3$ (the time to decline from peak by two and three magnitudes, respectively) and the rise time $\Delta t_0$ \citep[here, using Equation~16 from][]{Hachisu2006}. Eruption parameters computed via this approach are recorded in Table \ref{tab:Vtimings}.

\begin{table}
\caption{Key parameters of the V392\,Per eruption.\label{tab:Vtimings}}
\begin{center}
\begin{tabular}{lll}
\hline
Methodology: & Discovery at $T_\mathrm{max}$ & Plateau at $t_3$ \\ 
\hline
Discovery (MJD) & \multicolumn{2}{c}{58237.474} \\
Eruption:\ $T_0$ (MJD) & $58233^{+3}_{-2}$ & $58236.0\pm0.2$ \\
Maximum light:\ $T_\mathrm{max}$ (MJD) & 58237.474 & $58237.1\pm0.3$ \\
Rise time:\ $\Delta t_0$ / days & $4^{+2}_{-3}$     & $1.1\pm0.2$ \\
$t_2$ / days & $3.1\pm0.2$ & $2.0\pm0.1$ \\
$t_3$ / days & $8 ^{+1}_{-2}$ & $4.2\pm0.3$ \\
Plateau onset / days & $12\pm1$ & $5.2^{+0.9}_{-1.1}$ \\
Plateau duration / days & $3\pm2$ & $5\pm1$ \\ 
$m_{V,\mathrm{max}}$ / mag & $5.92^{-0.04}_{+0.3}$ & $5.51\mp0.09$ \\
$E\left(B-V\right)$ / mag & \multicolumn{2}{c}{$0.70^{+0.03}_{-0.02}$} \\
$M_{V,\mathit{Gaia}}$ / mag & $-9.0^{-0.6}_{+0.4}$ & $-9.4^{-0.4}_{+0.3}$ \\
$M_{V,\mathrm{MMRD}}$ / mag & $-8.5\mp0.2$ & $-8.8\mp0.2$\\
$d_\mathit{Gaia}$ / kpc & \multicolumn{2}{c}{$3.5^{+0.6}_{-0.5}$} \\
$d_\mathrm{MMRD}$ / kpc & $2.7\pm0.5$ & $2.7\pm0.3$ \\
\hline
\end{tabular}
\end{center}
\end{table}

However, the light curve follows the P-class morphology \citep*{Strope2010}, exhibiting a pseudo-plateau in the early evolution (see Figure \ref{fig:LCAll_log}), where the otherwise smooth, steep decline has a relatively flat interval, typically 3--6\,mag below peak, followed by another steep decline. If we assume that $T_\mathrm{max}$ coincides with the earliest $V$-band observation ($V=6.36$\,mag), then plateau onset occurs after a decline of only 2.1\,mag. Thus $t_3$ occurs during the plateau, interrupting the smooth decline, which leads to a relatively long rise time estimate ($\Delta t_0=4^{+2}_{-3}$\,days) for such a rapidly evolving eruption. Here, poor constraint of the eruption time leads to large uncertainties on all light curve derived parameters. 

As an alternative, we assumed that plateau onset coincides with $t_3$. In the \citet{Strope2010} sample of 19 P-class novae, only V4021\,Sgr entered a plateau earlier (2.4\,mag below peak; it also had the slowest decline of the sample). Fixing plateau onset at $t_3$ provides a conservative estimate of the time of maximum:\ if the plateau onset occurs later there would have been an earlier and brighter peak. The light curve evolution of V392\,Per is well described by a series of broken power-laws (see Figure \ref{fig:LCAll_log}), whose indices depend upon the assumed $T_0$. Hence, an iterative approach was used to fit the light curves (see Section~\ref{sec:LCfitting}) to determine $T_\mathrm{max}$ such that the plateau began at $t_3$. This leads to independent estimates of $T_0$, $\Delta t_0$, $t_2$, $t_3$, and  $T_\mathrm{max}=0.3\pm0.3$\,days pre-discovery (see Table~\ref{tab:Vtimings}). 

Regardless of the method employed, $t_2<4$\,days:\ a very fast eruption \citep{1964gano.book.....P}, and V392\,Per is one of the fastest evolving novae yet discovered. Based on the likelihood that maximum light was missed, the rapid evolution of the light curve, the $\gamma$-ray detection, and behaviour of similar P-class novae, we adopt these estimates throughout.

\subsection{Distance, Extinction, and Astrometry}\label{sec:reddening}

\citet{Stoyanov2020} measured radial velocities of diffuse interstellar bands and interstellar \ion{K}{i} in their V392\,Per spectra from 2018 May 1--2, deriving $E(B-V)=1.2\pm0.1$. \citet{Munari2020} compared the $(B-V)$ colour of V392\,Per shortly after peak with the expected intrinsic colour at maximum to derive $E(B-V)=0.72\pm0.06$. The \citet{Stoyanov2020} measurement was very early post-eruption and the ejecta may have added to the extinction column. 

The equivalent width of the interstellar sodium doublet absorption line is often used to determine reddening. However, the interstellar \ion{Na}{i}-D lines were saturated in our spectra. \citet{Stoyanov2020} also reported saturation of the Na doublet.

The astrometry of V392\,Per, as reported by {\it Gaia} EDR3 \citep{2016A&A...595A...1G,2021A&A...649A...1G} is $\alpha=4^\mathrm{h}43^\mathrm{m}21^\mathrm{s}\!.369814\pm0.04\,\mathrm{mas},\ \delta=47^\circ21^\prime25^{\prime\prime}\!\!.84112\pm0.03\,\mathrm{mas}$ (J2000). EDR3 reports a parallax measurement for V392\,Per of $\varpi=0.276\pm0.046$\,mas. Following \citet{2021AJ....161..147B}, this leads to a distance estimate of $d=3.5^{+0.6}_{-0.5}$\,kpc. Utilising the 3D dust maps of \citet{2019ApJ...887...93G}, we estimate the line of sight reddening toward V392\,Per to be $E\left(B-V\right)=0.70^{+0.03}_{-0.02}$. This follows the approach used by \citet{DarStar18}\footnote{The parallax reported by \citet{DarStar18} was actually that of RS\,Oph, although their reported distance did relate to V392\,Per.}, however, both the distance and reddening estimates are smaller due to advances between {\it Gaia} DR2 and EDR3. This reddening estimate is in agreement with that by \citet{Munari2020} and we adopt $E\left(B-V\right)=0.7$ throughout. As such, and utilising the plateau method (see Section~\ref{sec:Tnought}), we estimate a peak absolute magnitude $M_{V,\mathit{Gaia}}=-9.4^{-0.4}_{+0.3}$\,mag. This large {\it Gaia} distance and resulting luminous $M_V$ demonstrate that V\,392 Per is NOT a ``faint-fast" nova, like those commonly seen in M\,31 and in M\,87 \citep{Kasliwal2011,Shara2016}. Thus the use of the MMRD is justified to check on the {\it Gaia} distance. The `S-shaped' MMRD calibrated by \citet[see their Equation~15]{2020A&ARv..28....3D} for a nova with $t_2=2.0\pm0.1$  produces a consistent ($<2\sigma$) estimate of $M_{V,\mathrm{MMRD}}=-8.8\mp0.2$, and an MMRD distance estimate of $2.7\pm0.3$\,kpc (within 1.4$\sigma$ of the \textit{Gaia} distance).

\subsection{Photometry and light curve fitting}\label{sec:LCfitting}

Figure \ref{fig:LCAll_log} presents the $u'BVr'i'z'$ and \textit{Swift}/UVOT uvw1, uvm2, and uvw2 light curves for V392\,Per. The optical observations are shown with the same scale to aid comparison, and the $BVr'i'$ light curves include observations taken by AAVSO observers (all photometric data before the first Sun constraint: see Table \ref{tab:AAVSOobservers} for observer details). The colour evolution of the nova is also shown in Figure \ref{fig:LCAll_log}. The series of high-cadence $i'$-band photometry collected by LT and LCOGT is included and illustrates the high amplitude variations seen (see Section~\ref{sec:orbper}).

As shown in Figure \ref{fig:LCAll_log}, the optical and near-UV light curves of V392\,Per can be broadly replicated by a series of six broken power laws ($f\propto t^\alpha$) and a least-squares regression was employed to fit each light curve. Key parameters from the best fits are shown in Table \ref{tab:Fits_new}. In general, the light curve exhibits an initial decline from maximum before entering a quasi-plateau after $\sim5$\,days. The plateau continues for a further $\sim5$\,days after which the decline steepens further and the light curve follows three broken power laws as it approaches an approximately flat luminosity $\sim225$\,days post-eruption.

The onset, duration, and gradient of the plateau differs between the passbands; with a shallower gradient, later onset and longer duration with decreasing wavelength. Such plateaus have been proposed to be caused by a surviving, or reformed, accretion disk emerging from the optically thick photosphere as it recedes back toward the WD surface \citep{Hachisu2006,2018ApJ...857...68H}. The behaviour  here is compatible with a cooler outer disk emerging from the receding photosphere earlier than the inner hotter regions.

There is a potential light curve discontinuity during the first Sun constraint.  The $B$-band light curve is poorly sampled but appears continuous across the gap, but the $Vr'i'$ data point to a change in gradient during that Sun constraint, possibly hinting at a further plateau stage during the gap. Upon exiting the initial Sun constraint, the system had entered the nebular phase, with strong emission from [$\ion{O}{iii}$] 4363, 4959, and 5007\,\AA\ and \ion{He}{ii} 4686\,\AA\ present (see Section \ref{sec:SpecAnalysis}). These emission lines began to appear before the end of the Sun constraint and, due to their strengths, may have driven the changes observed in the $B$ and $V$ light curves.

The power-law indices in the initial decline in the $BVr'$-bands are $\alpha=-1.75\pm0.04$, $-1.76\pm0.07$, and $-1.78\pm0.8$, respectively. This appears in good agreement with the expected continuum from free-free emission from an optically thin plasma (also see Section~\ref{sec:shock_lc} for discussion about the nature of the early-decline light curve). Other than the initial power law and the plateau, we do not ascribe any physical meaning to the power laws. We simply utilise these as a tool to calibrate the optical spectroscopy (see Section \ref{sec:SpecAnalysis}). 

\begin{figure}
	\includegraphics[width=0.9\columnwidth]{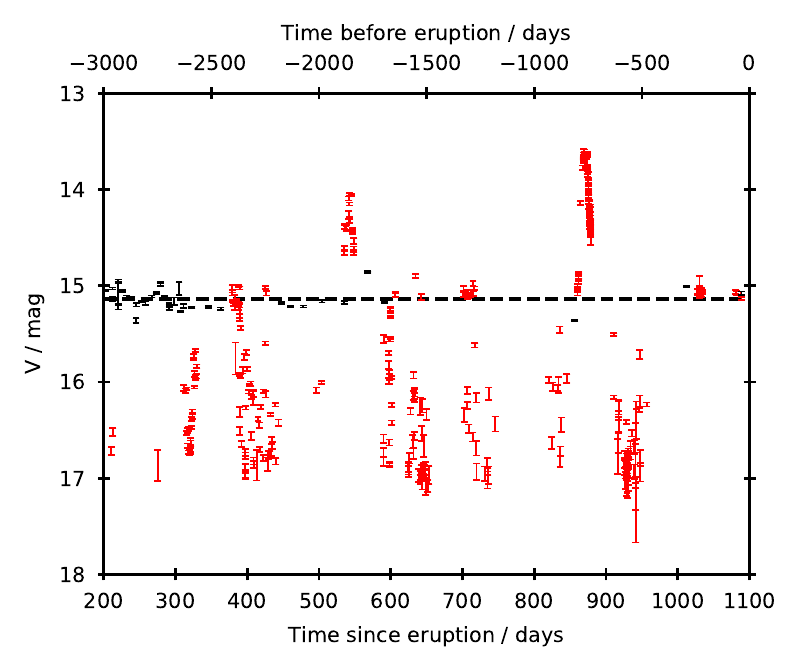}
    \caption{Comparison of post-nova (black) and pre-nova (red) $V$-band brightness of V392\,Per. The average post-nova magnitude $\bar{V}=15.2 \pm 0.1$\,mag is shown by the dashed line. The time-scale for the pre-eruption AAVSO data from March 2004 to March 2018, which covers around 3 times the duration of the post-nova observations, is shown by the top time axis.}
    \label{fig:post_vs_pre_nova}
\end{figure}

The light curves have remained broadly static at the post-nova luminosity since $\sim225$\,days post-eruption, with $\bar{V}=15.2\pm0.1$\,mag. This is substantially brighter than the long-term quiescent minimum of $\sim17$\,mag, and was referred to by \citet{Munari2020} as `sustained post-(eruption) brightening', see Figure~\ref{fig:post_vs_pre_nova}. 

\subsection{Spectral Energy Distribution}\label{sec:SED}
\begin{figure*}
\begin{center}
\includegraphics[width=0.8\textwidth]{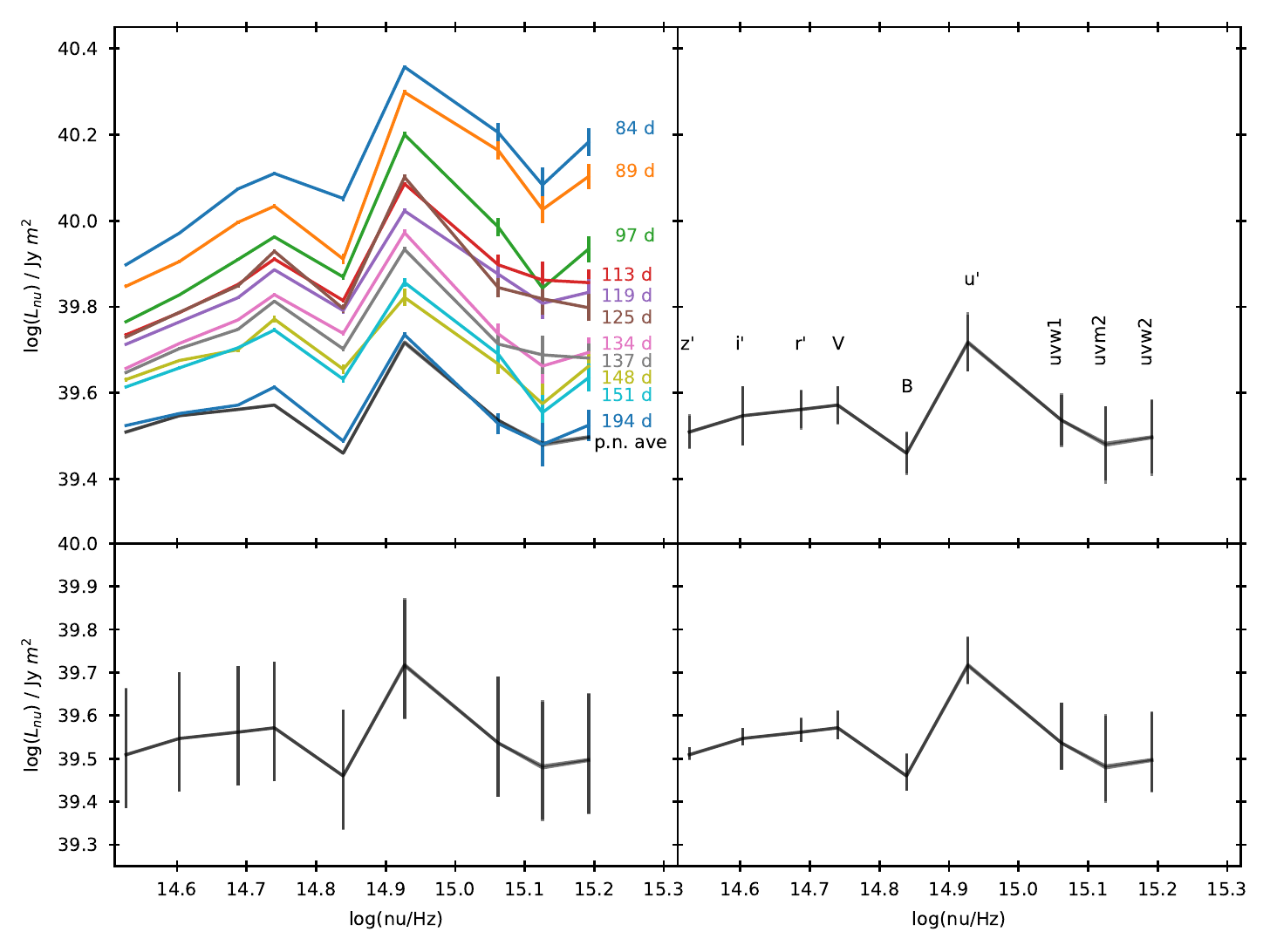}
\end{center}
\caption{Spectral energy distribution (SED) evolution of V392\,Per. {\bf Top left}: Evolution between days 84--194, and the average post-nova (p.n. ave) SED, from day 220 and beyond, is shown in black at the bottom. {\bf Top right}: Average post-nova SED, error bars indicate $1\sigma$ scattter. {\bf Bottom left}: Distance uncertainty, $d=3.5^{+0.6}_{-0.5}$\,kpc {\bf Bottom right}: Extinction uncertainties, $E\left(B-V\right)=0.70^{+0.03}_{-0.02}$.\label{fig:SEDaverage}}
\end{figure*}

The evolution of the spectral energy distribution (SED) of V392\,Per is shown in Figure \ref{fig:SEDaverage}. The SEDs are derived from dereddened photometry through LT $u'BVr'i'z'$, and \textit{Swift} uvw2, uvm2 and uvw1 filters. We have assumed the {\it Gaia} determined distance of 3.5\,kpc and $E\left(B-V\right)=0.7$\,mag. All plots include the SED averaged over the post-nova period, $t\geq223$\,days post-eruption. 

Since V392\,Per emerged from the first Sun constraint, the SED shape has remained broadly constant, with the overall luminosity fading toward the post-nova average (black line), although the overall SED slope has gradually decreased: the SED has become redder. From day 194 post-eruption, the SED luminosity has remained very close to the average post-nova value. From day 84, the SED shows a persistent $V$-band bump, which seems to be driven by [\ion{O}{iii}] 4959+5007\,\AA\ emission. We propose that the SED from the $u'$-band and bluer is that of an accretion disk (see Section \ref{sec:Disc}).

\subsection{Orbital Period}\label{sec:orbper}

The post-nova light curve of V392\,Per shows clear and significant variation, see Figure~\ref{fig:LCAll_log} ($i'$-band) and Figure~\ref{fig:post_vs_pre_nova}. There are three published periods for V392\,Per: 

\begin{figure*}
\begin{center}
\includegraphics[width=0.9\columnwidth]{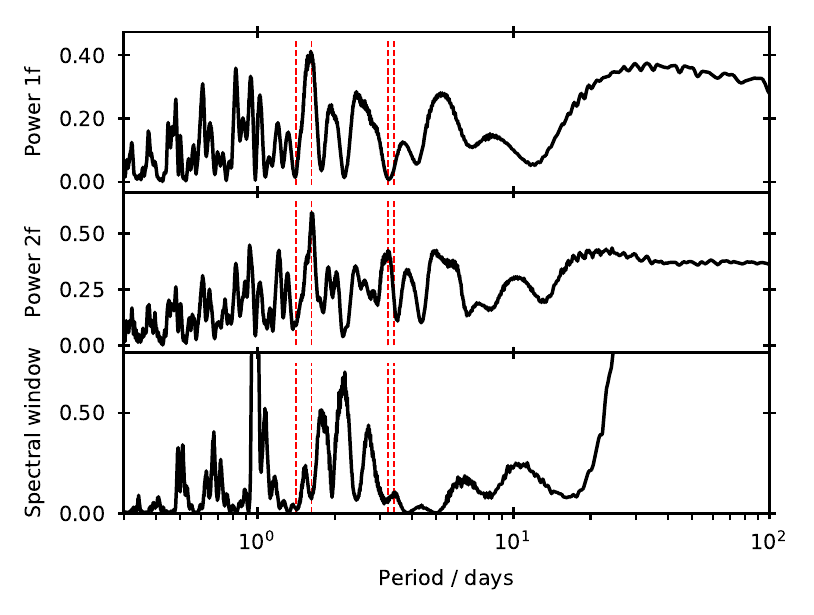}\hfill\includegraphics[width=0.9\columnwidth]{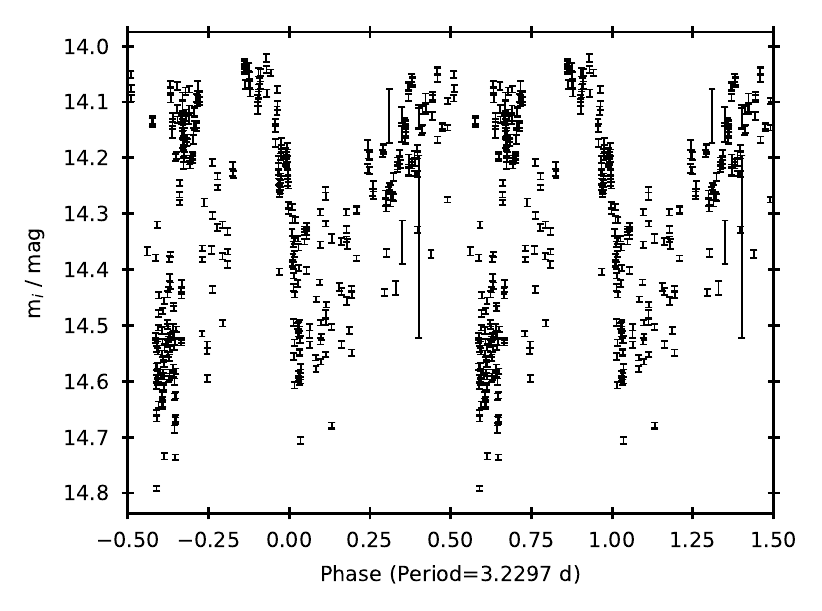}
\end{center}
\caption{\textbf{Top left:} Power spectrum of post-nova $i'$-band observations using a single Fourier sinusoidal term. {\bf Middle left:} As top left, but utilising two Fourier terms.  \textbf{Bottom left:} Window function power spectrum. In all three panels the vertical red dashed lines indicate the reported periods, from left to right: $P=1.4107$\,days \citep{Munari2020}; $P=1.62419\pm0.00069$\,d (this work); $P=3.21997 \pm 0.00039$\,days \citep{Schaefer2021}; $P=3.2297\pm0.0027$\,days \citep[this work; indistinguishable from][]{Schaefer2021}; and $P=3.4118$\,days \citep{Munari2020}. {\bf Right:} Post-eruption $i'$-band light curve, folded on best fit period of $P=3.2297\left(\pm0.0027\right)$\,days. \label{fig:powerspec}}
\end{figure*}

\citet{Schmidt2020} used Cousins $I$-band photometry collected over 78\,days between 2019 December 22 and 2020 March 9 \citep[effectively the same period as][]{Munari2020}. The data were detrended by subtracting the nightly mean magnitude. \citet{Schmidt2020} performed a discrete Fourier transform and Lomb-Scargle periodogram analysis, yielding a period of $P=0.06600 \,\pm\,0.00002$\,days. 

The \citet{Munari2020} period was calculated using ANS Collaboration $VRI$ data taken over 17 nights between 2019 December 30 and 2020 March 11. The Fourier power spectrum of these data revealed two significant, potentially linked peaks, $P=3.4118$\,days and $P=1.4107$\,days (both illustrated in Figure~\ref{fig:powerspec}). Consideration of the pre-nova photometry and derived system parameters led them to favour the longer period. \citeauthor{Munari2020} notes that their $R$ and $I$-band data show similar periodic modulation.

\citet{Schaefer2021} utilised 1150 TESS observations from 2019, with 28725 supplementary AAVSO observations from 2019--2021, and, following cleaning and detrending, employed a Fourier technique and folded light curve fitting to estimate $P=3.21997 \pm 0.00039$\,days, with an amplitude of 0.122 magnitudes. However, the TESS CCD scale ($21^{\prime\prime}$\,pixel$^{-1}$) would prohibit the disentanglement of signals from V392\,Per and the nearby, similar luminosity, field star (standard \#15; $9^{\prime\prime}$ distant). Some AAVSO observers were unable to separate these two sources as the nova faded. \cite*{Munari2020ATel} find the neighbour star shows no variability; our photometry of this source concurs.

We collected 423 high-cadence observations in the $i'$-band using LT and LCOGT between 2019 November 17 and December 2. These data show variation with amplitude up to $\sim \,0.7$\,mag over the course of a night, much greater than reported by \citeauthor{Munari2020} or \citeauthor{Schaefer2021}. The left panel of Figure~\ref{fig:powerspec} shows the Lomb-Scargle power spectrum for all our $i'$-band observations taken after day 252, during the roughly consistent brightness post-nova phase, and we also show the associated window function. The periods reported by \citet{Munari2020} and \citet{Schaefer2021} are indicated on the power spectrum. The strongest significant peak is found at $P=1.62419\pm0.00069$\,d, when utilising a single (sinusoidal) Fourier term; this is very close to half the \citet{Schaefer2021} value. Adding a second sinusoidal term reveals an additional peak at $P=3.2297\pm0.0027$\,days, very close to \citeauthor{Schaefer2021}'s. The right panel of Figure \ref{fig:powerspec} presents our $i'$-band data folded around $P=3.2297$\,days. Here, upon visual inspection, there does appear to be a plausible phase-folded light curve that is compatible with a double-dipping CV. However, the folded light curve appears `noisy', and we suggest that this may be due to different periodic, or other, activity from the system.

\subsection{Optical spectra}\label{sec:specresults}

\begin{figure*}
\begin{center}
\includegraphics[width=0.8\textwidth]{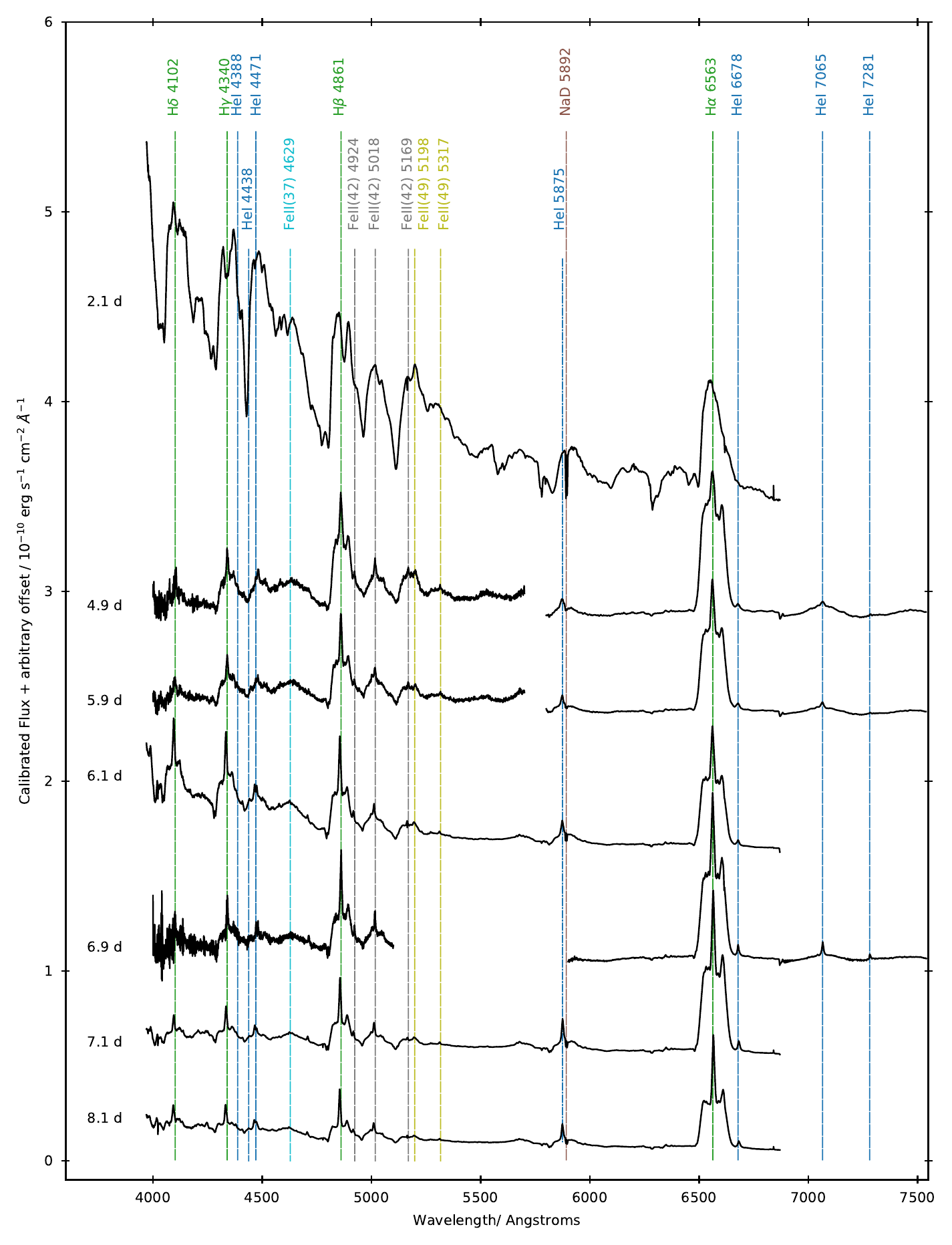}
\end{center}
\caption{Pre-first Sun constraint spectra of V392\,Per, from 2.1--8.1\,days post-eruption. Here we present flux calibrated (but offset) dereddened spectra from the LT (SPRAT and FRODOSpec) and the Hiltner 2.4\,m. These early spectra are becoming progressively less optically thick. Prominent spectral features are labelled. Spectra are available online.\label{fig:Calspecearly}}
\end{figure*}

\begin{figure*}
\begin{center}
\includegraphics[width=0.8\textwidth]{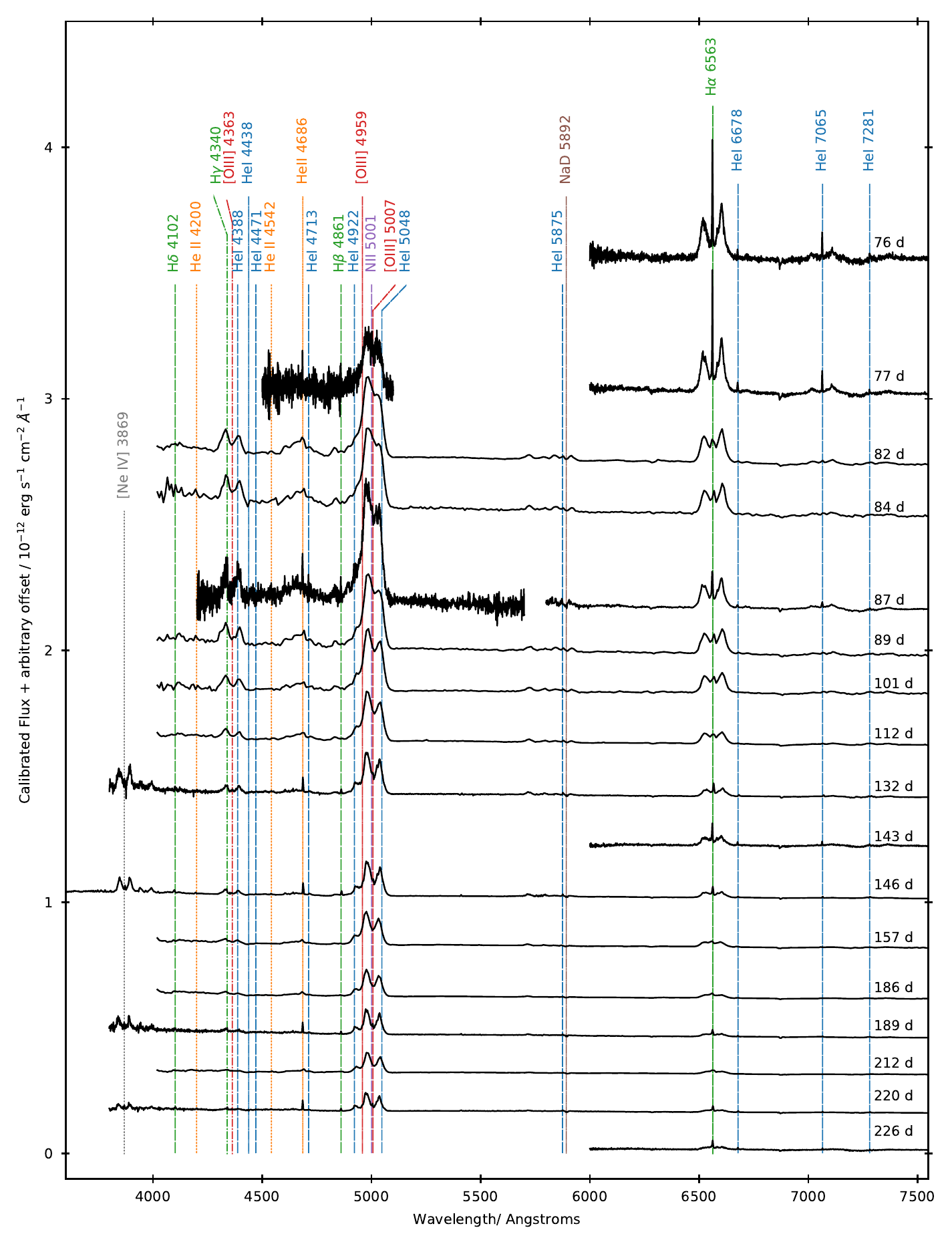}
\end{center}
   \caption{As Figure~\ref{fig:Calspecearly}, displaying spectra from 76--226\,days post-eruption and the nebular phase. Spectra are available online.}
    \label{fig:Calspecmid}
\end{figure*}

\begin{figure*}
\includegraphics[width=0.8\textwidth]{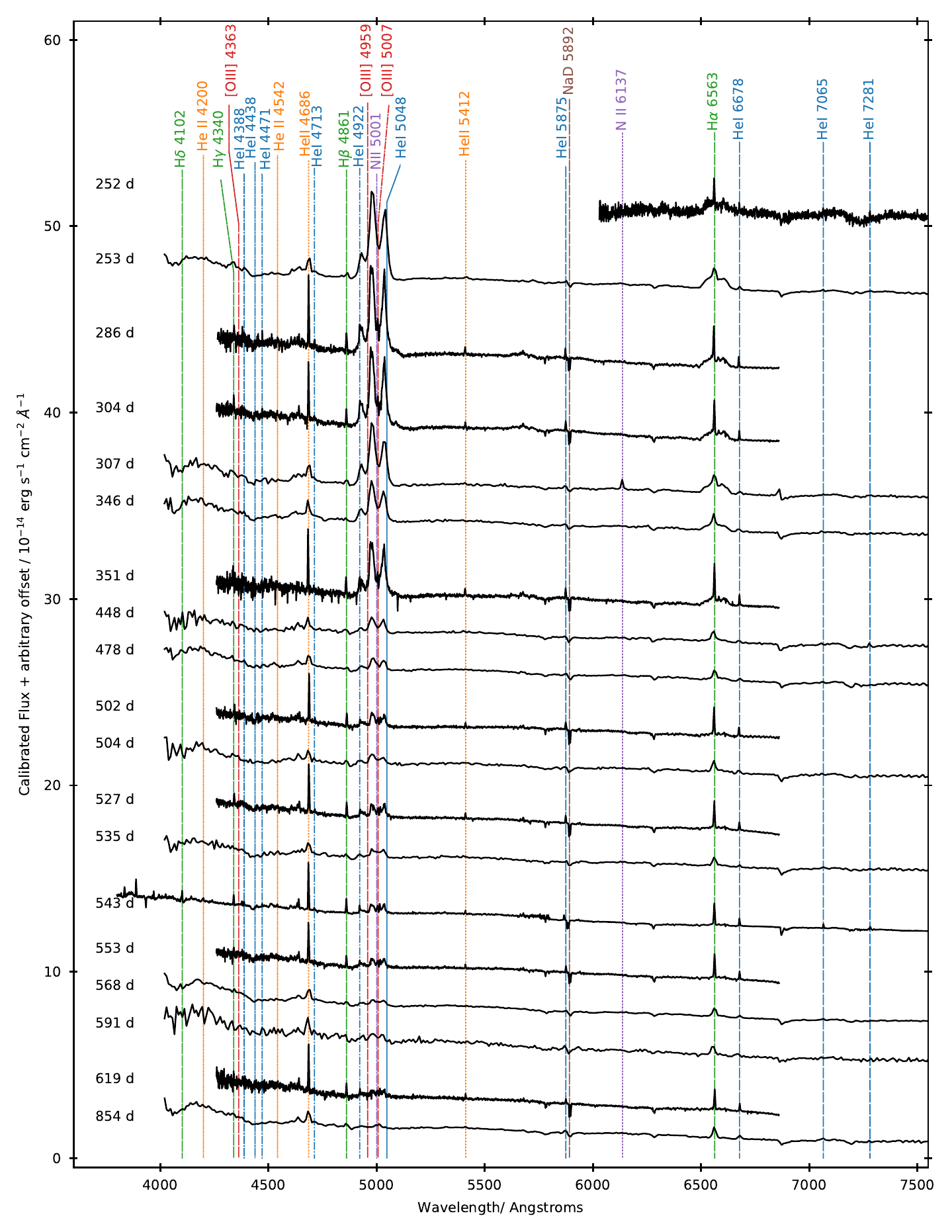}
    \caption[width=\textwidth]{As Figure~\ref{fig:Calspecearly}, displaying spectra from 252--854\,days post-eruption, showing the transition from the late-nebular phase to the post-nova phase. Spectra are available online.}
    \label{fig:Calspeclate}
\end{figure*}

Our LT and Hiltner 2.4\,m flux calibrated and dereddened spectra are shown in Figures~\ref{fig:Calspecearly}--\ref{fig:Calspeclate}. Those shown in Figure \ref{fig:Calspecearly} were taken before V392\,Per entered the initial Sun constraint, and cover the period of early spectral evolution, while the spectra were declining in optical thickness. The strongest features in the earliest spectrum ($2.1$\,days post-eruption; 1\,day post-maximum) are \ion{H}{i} Balmer series lines, with H$\alpha$--H$\delta$ exhibiting broad, optically thick, P\,Cygni profiles. All lines attributable to the eruption have P\,Cygni shapes. We see strong lines from \ion{He}{i} 4471\,\AA, in particular, and from \ion{He}{i} 4388\,\AA\ and 4438\,\AA. Lines from \ion{Fe}{ii} multiplet 42 are dominant features. A broad \ion{Na}{i}-D profile is punctuated by saturated interstellar absorption lines (see Section~\ref{sec:reddening}). In the second spectrum ($4.9$\,days post-eruption), we also see \ion{He}{i} 6678\,\AA, 7065\,\AA, and 7281\,\AA. In addition, lines from \ion{O}{i} 7774\,\AA, 8227\,\AA, and 8446\,\AA\ were present (but not shown in Figure~\ref{fig:Calspecearly}). All spectra before the initial Sun constraint exhibit these lines, but their intensity and optical depth diminishes over this first week of evolution, and the line profiles evolve from P\,Cygni profiles to triple-peaked structures \citep[also see][]{2018ATel11601....1D,2018ATel11617....1M,2018ATel11605....1T,Wagner2018}. Based on the spectral morphology, we would place this eruption in the \ion{Fe}{ii} taxonomic class, although the inferred ejecta velocities are higher than normally seen in spectra from this class.

Once V392\,Per became visible following the first Sun constraint (76\,days post-eruption), the spectra had transitioned to the nebular phase \citep[see Figure \ref{fig:Calspecmid};][]{2018ATel11846....1D}. We see the continued presence of Balmer and \ion{He}{i} emission; however, the spectra are dominated by nebular [\ion{O}{iii}] 4959+5007\,\AA\ and auroral [\ion{O}{iii}] 4363\,\AA\ lines, with \ion{He}{ii} emission (particularly at 4200\,\AA\ and 4686\,\AA) now also present. As reported by \citet{2018ATel11872....1D}, the [\ion{O}{iii}] 5007\,\AA\ line rivals H$\alpha$ in brightness, and [\ion{O}{iii}] 4363\,\AA\ and \ion{He}{ii} 4686\,\AA\ are stronger than H$\beta$. The forbidden lines are double-peaked, whereas the \ion{H}{i} and \ion{He}{i} lines retain the triple-peaked structure, and the widths of the forbidden lines are consistent with those of \ion{H}{i}. The final set of spectra are shown in Figure~\ref{fig:Calspeclate}. Here we witness the decline of the nebular emission and the transition to the post-nova spectrum. Emission from the [\ion{O}{iii}] lines fades relative to that from \ion{He}{ii} 4686\,\AA\ and \ion{H}{i}.

As first reported by \citet{Munari2018}, we also see evidence for [\ion{Ne}{iii}] 3869\,\AA\ in the Hiltner and LBT spectra taken on days 132, 146, 189 and 220 post eruption. However, we do not see evidence for [\ion{Ne}{iv}] 4715\,\AA. This line might blend with the \ion{He}{ii} 4686\,\AA\ profile, but we link the structure seen at $\sim\pm2000$\,km\,s$^{-1}$ around \ion{He}{ii} to a contribution from the ejecta (see Section~\ref{sec:HeII}).

\subsection{\textit{Swift} X-ray and UV observations}

{\it Swift} observations commenced as soon as V392\,Per emerged from the first Sun constraint on 2018 July 20 \citep{2018ATel11905....1D}. The {\it Swift}/UVOT photometry is shown in Figure~\ref{fig:LCAll_log}. Although starting much later, the near-UV light curves match the late decline and approximately flat post-nova phases seen in the optical. There is a slight upward trend in the near-UV brightness during the post-nova phase. The system is consistently fainter through the uvm2 filter (which lies between the uvw1 and uvw2 filters, and samples the 2175\,\AA\ `bump' in the interstellar extinction curve), suggestive of high extinction.

\begin{figure*}
\begin{center}
\includegraphics[width=0.31\textwidth]{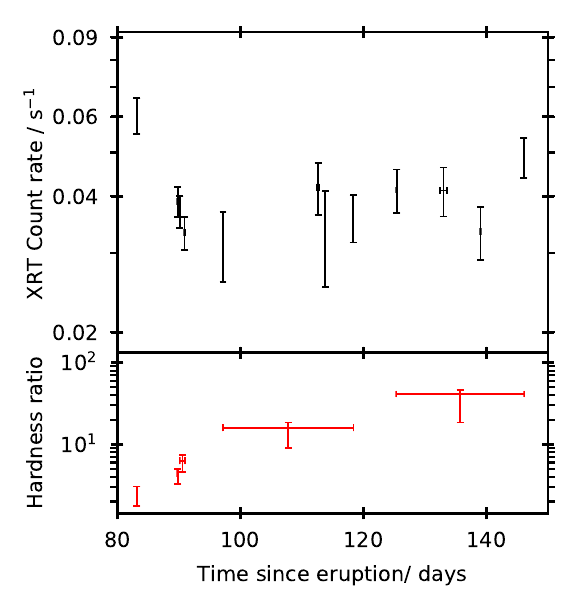}\hfill 
\includegraphics[width=0.31\textwidth]{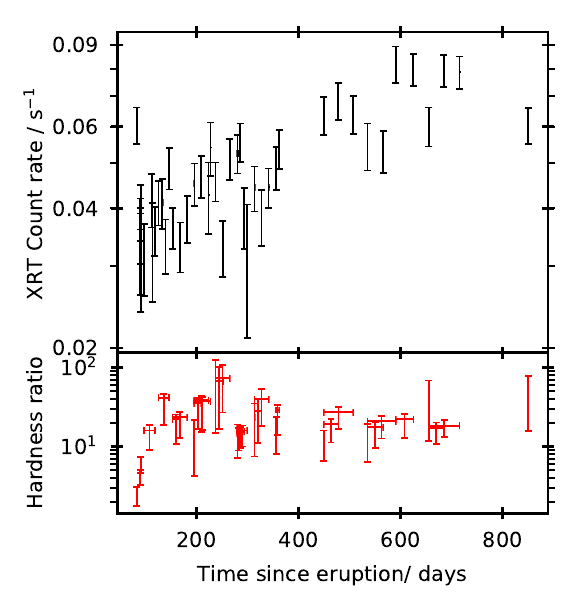}\hfill 
\includegraphics[width=0.37\textwidth]{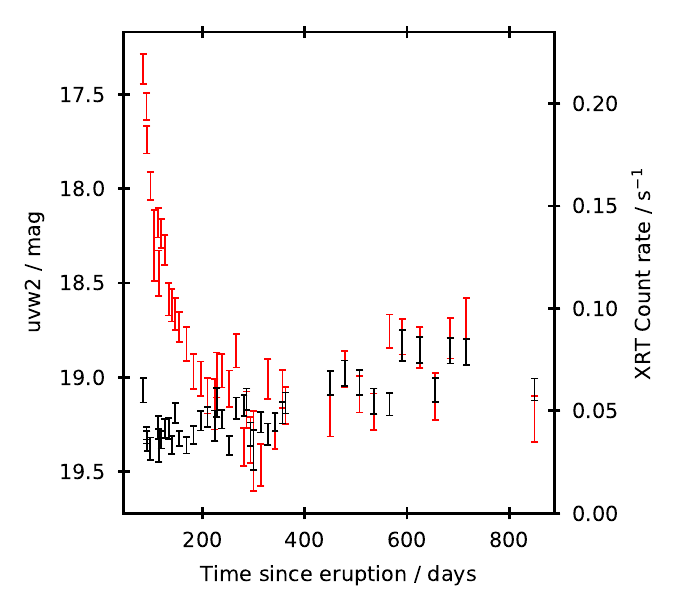} 
\end{center}
\caption{{\bf Left and centre:} {\it Swift}/XRT observations of V392\,Per. The upper panel shows the count rate, the lower panel shows the hardness ratio: counts(0.8--10\,keV)\ /\ counts(0.3--0.8\,keV). \textbf{Right:} {\it Swift}/XRT count rate (black) compared to the {\it Swift}/UVOT uvw2 photometry (red).\label{fig:SwiftXRTCounts_new}\label{fig:XRT_uvw2_new}}
\end{figure*}

The {\it Swift}/XRT light curve is presented in the left-hand panel of Figure~\ref{fig:SwiftXRTCounts_new}. The plot at the top (black) shows the XRT count rate. A rapid decline in counts is seen from days 83--97, after which the counts remain approximately flat until entry into the second Sun-constraint. Upon exiting the second constraint, the XRT counts remained slightly elevated. The X-ray HR [defined as counts (0.8--10\,keV) / counts (0.3--0.8\,keV)] is shown in the bottom panel (red). The HR is approximately constant (although slightly decreasing) from day 112 onward. However, the HR at day 83 is clearly lower (softer), and between days 83--97 there is a gradual hardening. Here we propose that the softer emission seen between days 83--97 is the tail of the super-soft source (SSS) phase of V392\,Per.

The right panel of Figure~\ref{fig:XRT_uvw2_new} presents a comparison between the \textit{Swift}/UVOT uvw2 lightcurve and the XRT light curve; here the final decline in the near-UV is particularly evident. From $\sim300$\,days post-eruption, the XRT count rate and uvw2 photometry appear roughly correlated. This suggests that the post-nova near-UV and X-ray emission have a similar origin.

\section{Spectral analysis}\label{sec:SpecAnalysis}

\subsection{Balmer lines}\label{sec:Balmerlines}

\begin{figure*}
\begin{center}
\includegraphics[width=0.8\textwidth]{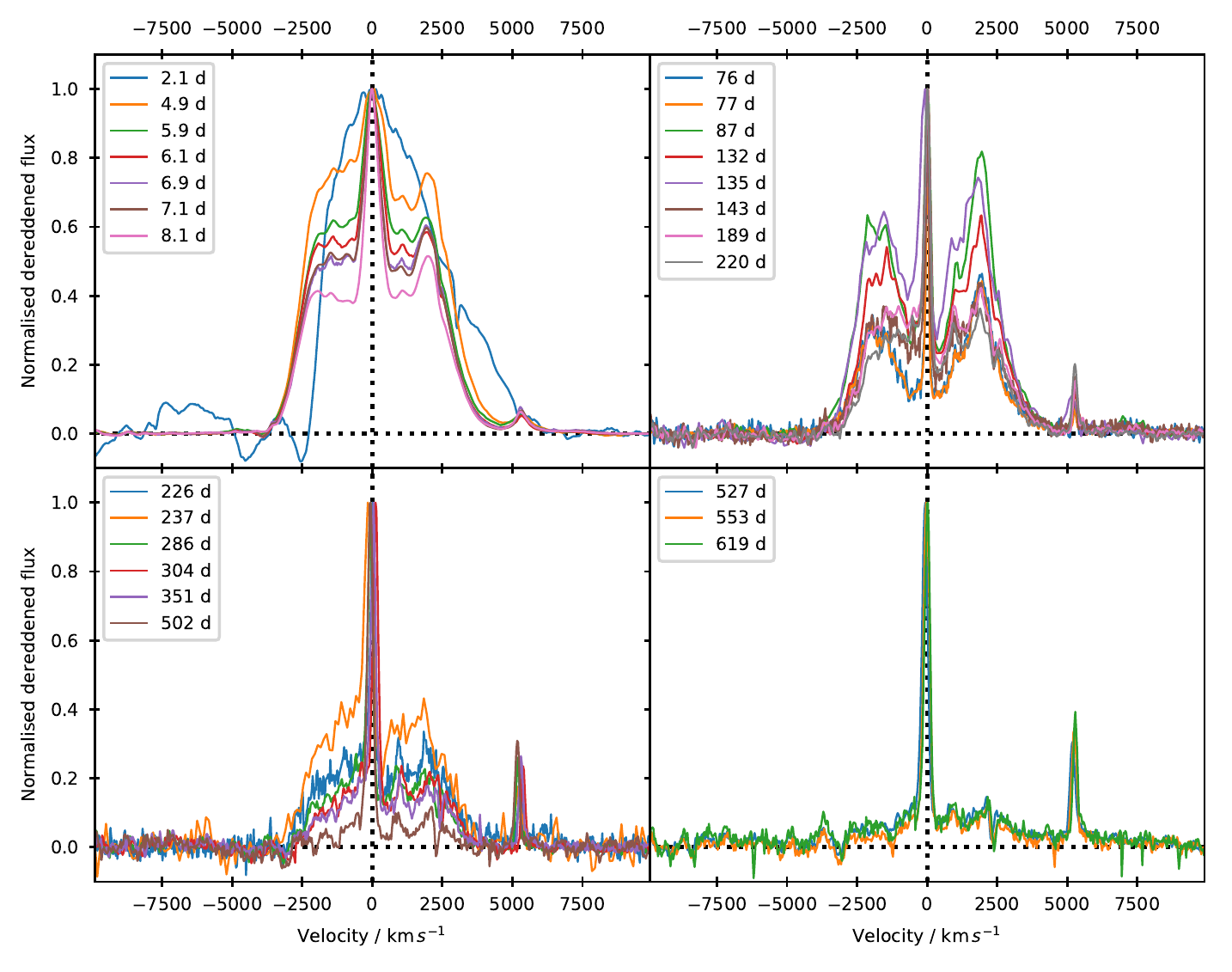}
\end{center}
\caption{High resolution H$\alpha$ line profiles from Hiltner OSMOS and FRODOSpec spectra. The spectra have been normalised to the flux of the H$\alpha$ central peak. The horizontal dotted lines indicate a normalised dereddened flux value of 0. The vertical dotted line shows the rest wavelength of H$\alpha$.\label{fig:Ha_hireslineprofiles_exc252}}
\end{figure*}

High resolution H$\alpha$ line profiles are shown in Figure~\ref{fig:Ha_hireslineprofiles_exc252}, which presents Hiltner 2.4\,m and the LT/FRODOSpec data. Our earliest spectrum (2.1\,days post-eruption; Figure~\ref{fig:Ha_hireslineprofiles_exc252} top-left) reveals a single-peaked, optically thick, broad and asymmetric H$\alpha$ line, with P\,Cygni absorptions at $-2560\, \mathrm{km}\,\mathrm{s}^{-1}$ and $-4550\,\mathrm{km}\,\mathrm{s}^{-1}$, and emission on the red side out to $\sim5000\, \mathrm{km}\,\mathrm{s}^{-1}$. By the next spectrum (4.9\,days post-eruption) the profile has developed a stronger and narrower central peak, and additional secondary peaks have started to develop at $\sim\!\pm2500\, \mathrm{km}\,\mathrm{s}^{-1}$, the P\,Cygni absorption elements have weakened and possibly merged (see Section~\ref{sec:Pcygni}), and the high velocity redward wing has gone.  Over the next week, as V392\,Per approached the first Sun constraint, the central peak increased in intensity relative to the red- and blue-shifted secondary peaks, with the redward peak more sharply defined than its blue counterpart.

The subsequent spectrum was taken once V392\,Per emerged from the first Sun constraint, on day 76 (Figure~\ref{fig:Ha_hireslineprofiles_exc252} top-right). By this time the system had evolved to the nebular phase and the H$\alpha$ line profile transitioned to a clear three-peaked structure with a bright, very narrow, central peak with measured FWHM of 57\,km\,s$^{-1}$. The velocity structure of the outer peaks are symmetric, but the redward peak is brighter. This morphology persisted until 220\,days post-eruption, but the outer peak amplitude continued to weaken relative to the central peak. 

From day 226 post-eruption (Figure~\ref{fig:Ha_hireslineprofiles_exc252} bottom-left), the amplitude of the central peak began to dominate and emission from the outer peaks began to wane more rapidly at higher velocities: the outer peaks appear to move inward toward the central peak -- most likely an effect of decreasing emissivity as the ejecta thin. The fastest moving ejecta thin the fastest. As the outer peak flux decreased, nearby lines became more prominent, e.g., \ion{He}{i} 6678\,\AA\ ($+5165\,\mathrm{km}\,\mathrm{s}^{-1}$). Between days 351--448 (see Figure~\ref{fig:Calspeclate}) all high velocity elements had disappeared, leaving just the narrow central line (Figure~\ref{fig:Ha_hireslineprofiles_exc252} bottom-right) --- the post-nova profile. 

\begin{figure*}
\begin{center}
\includegraphics[width=0.8\textwidth]{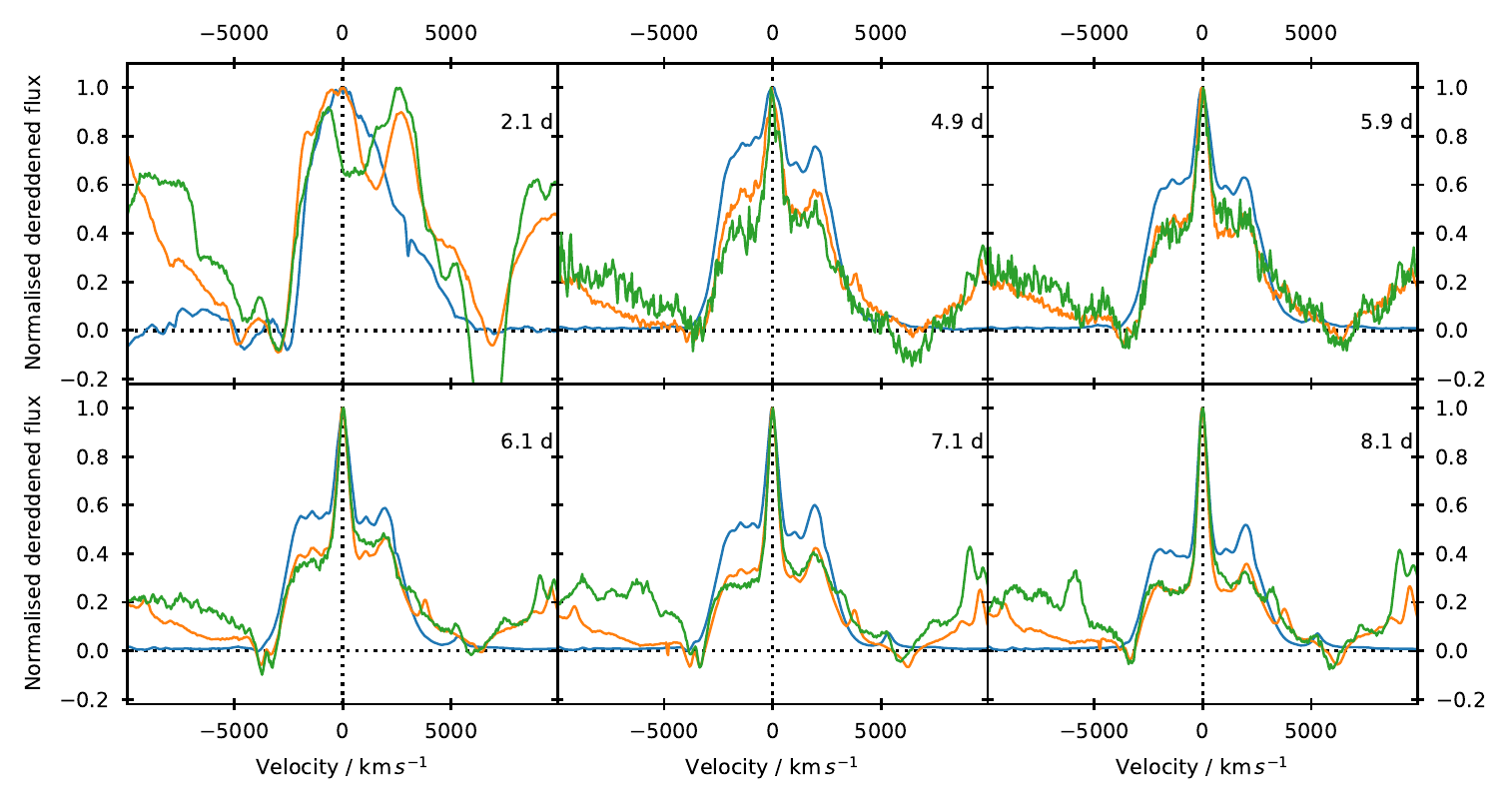}
\end{center}
\caption{Comparison of H$\alpha$ (blue), H$\beta$ (orange), and H$\gamma$ (green) line profiles for pre-first Sun constraint epochs. Spectra have been normalised to the central peak flux.\label{fig:Balmercomb}}
\end{figure*}

In Figure~\ref{fig:Balmercomb} we compare the H$\alpha$ profile with those of H$\beta$ and H$\gamma$. As expected, the Balmer line profiles evolve broadly similarly. The P\,Cygni absorptions persist for longer in H$\beta$ and H$\gamma$, possibly up to day 7.1 (see Section~\ref{sec:Pcygni}). The central peak is stronger relative to the outer peaks in H$\beta$ and H$\gamma$ than in H$\alpha$. This suggests there may be different recombination conditions in different ejecta components or stronger self-absorption in those components -- both indicative of a complex geometry. During the nebular phase, H$\beta$ and H$\gamma$ are severely blended with, and dominated by, the nebular and auroral [\ion{O}{iii}] lines, respectively. By the post-nova phase all three Balmer lines simply show a very narrow peak.

The H$\alpha$ profile remains isolated from other lines and is one of the strongest lines in the spectra at all times, whereas other bright lines, e.g., other \ion{H}{i}, \ion{He}{i-ii}, and [\ion{O}{iii}], were often severely blended. Thus, to measure the flux of the emission lines we used a triple-Gaussian profile, modelled around the H$\alpha$ profile at each epoch, to estimate line fluxes and, where necessary, de-blend lines. Line fluxes are tabulated in Tables~\ref{tab:Fluxes_all} and \ref{tab:Fluxes_all2}.

\begin{figure*}
\begin{center}
\includegraphics[width=0.9\columnwidth]{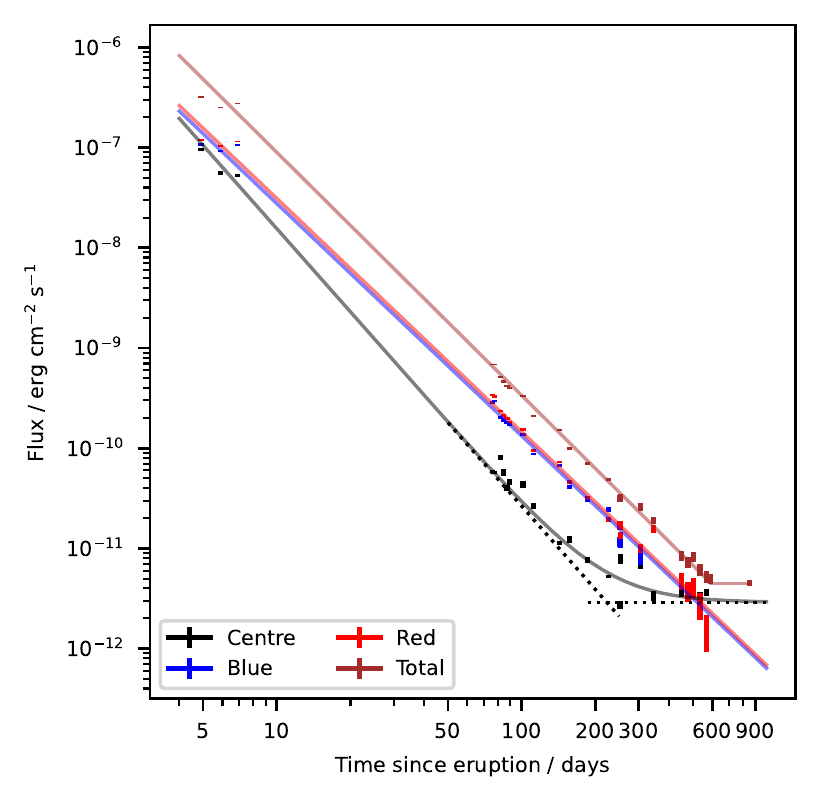}\hfill
\includegraphics[width=0.9\columnwidth]{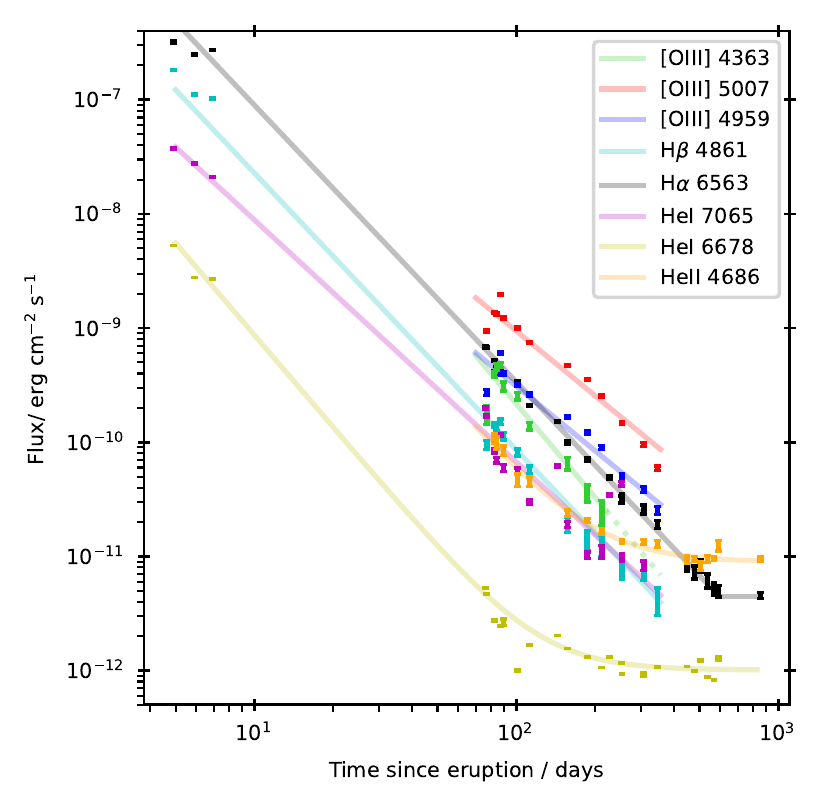}
\end{center}
\caption{\textbf{Left:} H$\alpha$ flux evolution from SPRAT and FRODOSpec spectra. The fluxes of the central, blueward and redward components are shown in black, blue and red, respectively. The total flux is shown in brown. \textbf{Right:} Flux evolution of prominent lines in the V392\,Per spectra. Data for [O\,III] 4959\,\AA, [O\,III] 5007\,\AA, and H$\beta$ are only shown up to day 346. The H$\alpha$ flux from day 346 onward only includes the central and redward component, and for day 591 onward only  the central component. Power law fits to these data are indicated by the solid lines, dotted lines are extrapolations of fits beyond the available data.\label{fig:Hafluxev}\label{fig:Fluxes350d_newflux}}
\end{figure*}

The flux evolution of the H$\alpha$ line profile is shown in Figure~\ref{fig:Fluxes350d_newflux} (left). Although the high amplitude of the central peak had appeared to dominate from day 226, here we see that the integrated emission from the outer peaks (blue and red) dominates the overall line flux (brown) until day $\sim\!600$. The decline of the outer peak fluxes are well described by a power-law with index $-2.32\pm 0.04$. The decay of the central peak (black) is steeper than the outer peaks and power-law-like until day $\sim\!100$, however, from this point the central peak tends towards a constant flux of $(2.90 \pm 0.42)\times 10^{-12}$\,erg\,s$^{-1}$\,cm$^{-2}$. Here we suggest that there are two system components contributing to the H$\alpha$ line, a triple-peaked ejecta profile, that decays as a power-law, and a constant single narrow peaked contribution from the central system. While the H$\beta$ line is strong and isolated during the early spectra, once in the nebular phase the line is severely blended and dominated by [\ion{O}{iii}] 4959+5007\,\AA. As such, we can only reliably estimate the H$\beta$ flux until day $\sim\!350$. As shown in Figure~\ref{fig:Fluxes350d_newflux}, the H$\beta$ flux declines following a similar power-law to H$\alpha$.

\subsection{Multiple ejections?}\label{sec:Pcygni}

\begin{figure*}
\begin{center}
\includegraphics[width=0.8\textwidth]{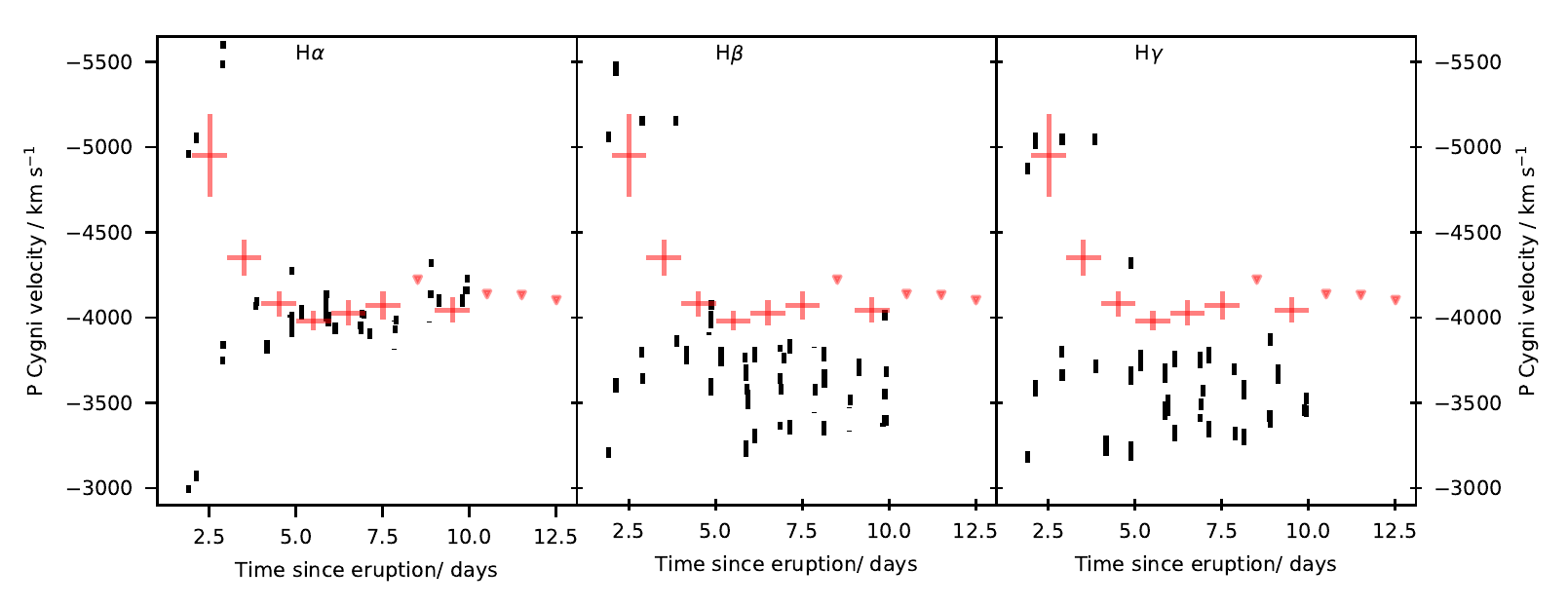}
\end{center}
\caption{Velocity evolution of the Balmer line (left:\ H$\alpha$, centre:\ H$\beta$, right:\ H$\gamma$) P\,Cygni absorption features (black). The red data points show the Fermi-LAT light curve for comparison, the red arrowheads are 95th percentile upper limits.}
    \label{fig:PCygvel}
\end{figure*}

In Figure~\ref{fig:PCygvel} we show the velocity evolution of detected P\,Cygni absorption components from the Balmer lines. For illustrative purposes, we also show the Fermi-LAT $\gamma$-ray light curve and 95th percentile upper limits \citep{2021ARA&A..59..391C,2022arXiv220110644A}. Here we also utilise the ARAS spectra, some of which included very high resolution data for the Balmer lines. The first H$\alpha$ P\,Cygni measurement is 1.9\,days post-eruption and yielded two components $\sim-3000$\,km\,s$^{-1}$ and $\sim-5000$\,km\,s$^{-1}$. The H$\alpha$ P\,Cygni absorptions appear to shift further blueward over the next three spectra. By day 3.85, the H$\alpha$ profile only revealed a single P\,Cygni absorption at $\sim-4000$\,km\,s$^{-1}$, with subsequent measurements showing similar P\,Cygni velocities. Most of the later H$\alpha$ line profiles only contain a single P\,Cygni absorption, but a higher resolution spectrum taken $7.8$\,days post-eruption indicated that the absorption contained sub-structure with similar nearby minima of $\sim-3800$\,km\,s$^{-1}$ and $\sim-4000$\,km\,s$^{-1}$. The overall structure of the Balmer lines is complex and this led to systematic difficulties in the P\,Cygni measurement, and as such the scatter seen in the H$\alpha$ measures from day 5 onward, is indicative of the associated systematic errors.

The first Fermi-LAT $\gamma$-ray detection occurred 2.5\,days post-eruption (one day post-discovery), the second detection a day later had less than half the flux of the first detection, and the $\gamma$-ray flux had almost halved again by day 4.5. During this period of rapid $\gamma$-ray fading, we observe the apparent merger of two P\,Cygni absorptions, leaving a single absorption line with a velocity intermediate to the two. This scenario is consistent with that reported in \citet{Aydi2020,Aydi2020ApJ} for V906 Car and other CNe, where two constituents of a multi-component ejecta merge, with the associated shocks driving $\gamma$-ray emission. Here, we propose that an initial $\sim3000$\,km\,s$^{-1}$ ejection merges with a subsequent $\sim5000$\,km\,s$^{-1}$ ejection, leaving a single component travelling at $\sim4000$\,km\,s$^{-1}$. With this merger seemingly occurring $2.5\pm0.5$ days after the initial eruption, the second ejection would have occurred at $1.0\pm0.3$ days post-eruption, i.e.\ around the time of optical maximum. Assuming that kinetic energy is largely conserved during the merger, the second ejection could have a mass up to 80\% that of the first.

The equivalent H$\beta$ and H$\gamma$ data appear richer, both showing evidence for the initial merger that drove the strong $\gamma$-ray peak. From day 5, there remains evidence for two absorption features at lower velocities ($\sim-3750$\,km\,s$^{-1}$ and $\sim-3250$\,km\,s$^{-1}$). These are markedly lower than P\,Cygni velocities seen in the H$\alpha$ profile during the same epochs, but we note that the H$\alpha$ line was already transitioning to an optically thin emission profile at this time. These data are admittedly noisy, but they hint at a second or on-going merger event; which may be driving the flat $\gamma$-ray emission during this time. The spectral coverage ends at day 10, but in this final spectrum there is a hint of a single, merged, P\,Cygni at $\sim-3550$\,km\,s$^{-1}$. We note that this corresponds to the final Fermi-LAT detection, although V392\,Per remained visible to Fermi beyond this time.

\subsection{He \textsc{i} 6678 \AA\ and 7065 \AA}

The \ion{He}{i} 6678\,\AA\ profiles were fitted simultaneously with the H$\alpha$, and are shown in Figures \ref{fig:Ha_hireslineprofiles_exc252} and \ref{fig:Balmercomb} at $\sim 5250$\,km\,s$^{-1}$. The flux evolution of \ion{He}{i} 6678\,\AA\ is shown in the right panel of Figure~\ref{fig:Fluxes350d_newflux}. Due to its proximity to H$\alpha$, only the central peak of the \ion{He}{i} 6678\,\AA\ was measured, and emission peaked 4.9\,days post-eruption at $\left(5.30\pm0.01\right)\times 10^{-9}$\,erg\,s$^{-1}$\,cm$^{-2}$. The \ion{He}{i} 6678\,\AA\ flux follows a power law with index $-2.69\pm 0.16$, declining to a plateau of around $(1.01\pm 0.03)\times10^{-12}$\,erg\,s$^{-1}$\,cm$^{-2}$, from day 253 post-eruption.

As \ion{He}{i} 7065\,\AA\ is isolated from other strong lines, the profile was modelled using a three component Gaussian, as for H$\alpha$. The evolution of the total flux of \ion{He}{i} 7065\,\AA\ is shown in the right panel of Figure~\ref{fig:Fluxes350d_newflux}. The first flux measurement of \ion{He}{i} 7065\,\AA\ was $\left(3.74\pm0.04\right)\times 10^{-8}$\,erg\,s$^{-1}$\,cm$^{-2}$, 4.9\,days post-eruption. The \ion{He}{i} 7065\,\AA\ broadly follows a power law with index $-2.12\pm 0.05$, but there is no evidence for a flux plateau from the central component, as was seen in H$\alpha$ and \ion{He}{i} 6678\,\AA.

\subsection{He \textsc{ii} 4686 \AA}\label{sec:HeII}

\begin{figure*}
\begin{center}
\includegraphics[width=0.9\columnwidth]{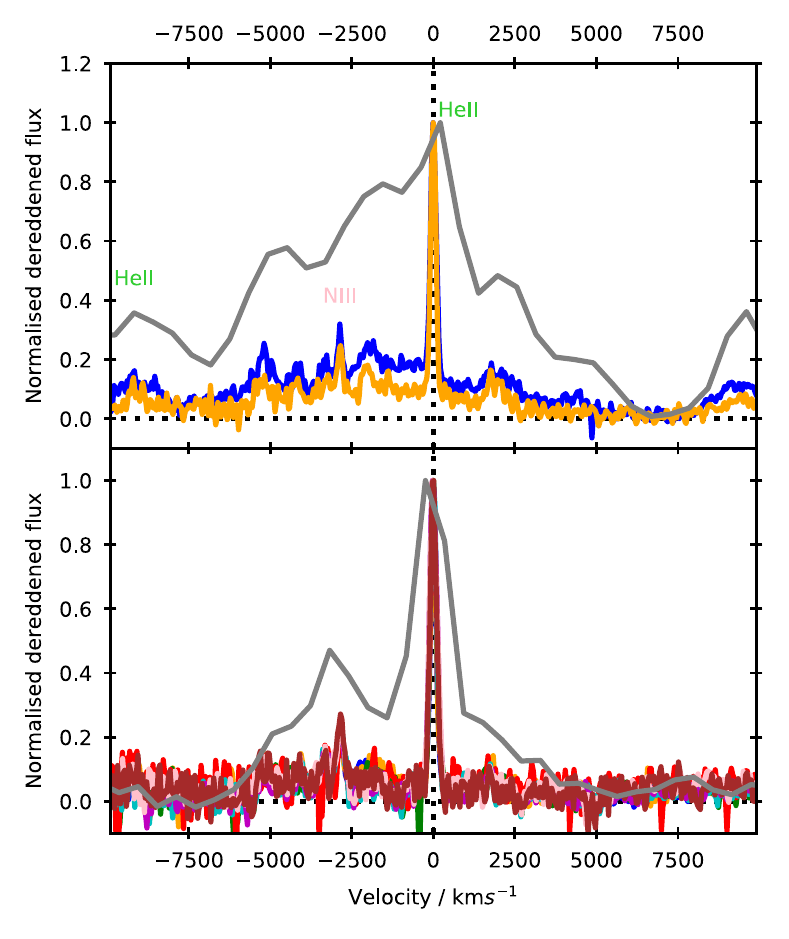}\hfill\includegraphics[width=0.9\columnwidth]{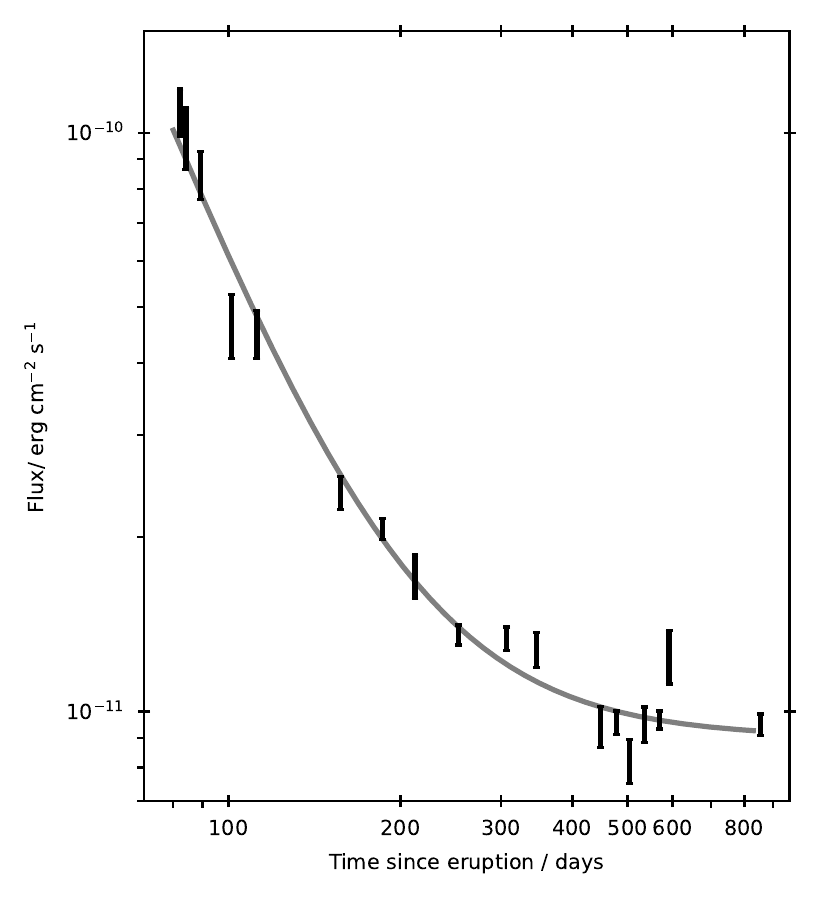}
\end{center}
\caption{\textbf{Left panels:} \ion{He}{ii} 4686\,\AA\ line profile evolution: high resolution profiles are compared with time-averaged low resolution data (grey) from the same time interval. \textbf{Top:} spectra from 83--212\,days post-eruption, high resolution data from 132 (blue) and 189 (orange) days post-eruption. \textbf{Bottom:} spectra from 253--854 days post-eruption. \textbf{Right:} \ion{He}{ii} 4686\,\AA\ flux evolution, where the solid line indicates a power-law plus plateau fit to the data.\label{fig:HeIIprofiles}\label{fig:HeIIflux}}
\end{figure*}

\ion{He}{ii} 4686\,\AA\ normalized line profiles are shown in Figure \ref{fig:HeIIprofiles} (left). The top panel shows profiles from the nebular phase following emergence from the initial Sun constraint 83--212\,days post-eruption. During this stage, the \ion{He}{ii} emission strengthened relative to neighbouring permitted lines. The low resolution spectra are suggestive of broad \ion{He}{ii} emission associated with the nova ejecta, due to blending from neighbouring lines. However, the higher resolution data indicate that the \ion{He}{ii} emission is dominated by a narrow central peak, with hints of a faint, broad, contribution from the ejecta ($\pm2300$\,km\,s$^{-1}$) in spectra from days 132 and 189.  The \ion{He}{ii} profiles toward the end of the nebular phase and throughout the post-nova period, show only the narrow central peak. 

Given the simplicity of the \ion{He}{ii} profile, we fitted the line using a single Gaussian. The flux evolution of \ion{He}{ii} 4686\,\AA\ is shown in Figure~\ref{fig:HeIIflux} (right). There is no significant detection of \ion{He}{ii} before the first Sun constraint. The first clear detection of \ion{He}{ii} occurs after emergence from this Sun constraint, on day 82. Here, the flux is ($1.09\pm 0.10) \times 10^{-10}$\,erg\,s$^{-1}$\,cm$^{-2}$. The evolution of the line flux is best described by the combination of a  power law of index $-2.54\pm 0.16$ and a plateau of around $(9.02 \pm 0.37) \times 10^{-12}$\,erg\,s$^{-1}$\,cm$^{-2}$. The flux evolution is compared with that of other species in Figures~\ref{fig:Fluxes350d_newflux} (right), shown in Figure~\ref{fig:HeIIflux}, and discussed further in Section \ref{sec:Disc}.

\subsection{Nebular [O \textsc{iii}] 4959+5007 \AA}

\begin{figure*}
\begin{center}
\includegraphics[width=0.8\textwidth]{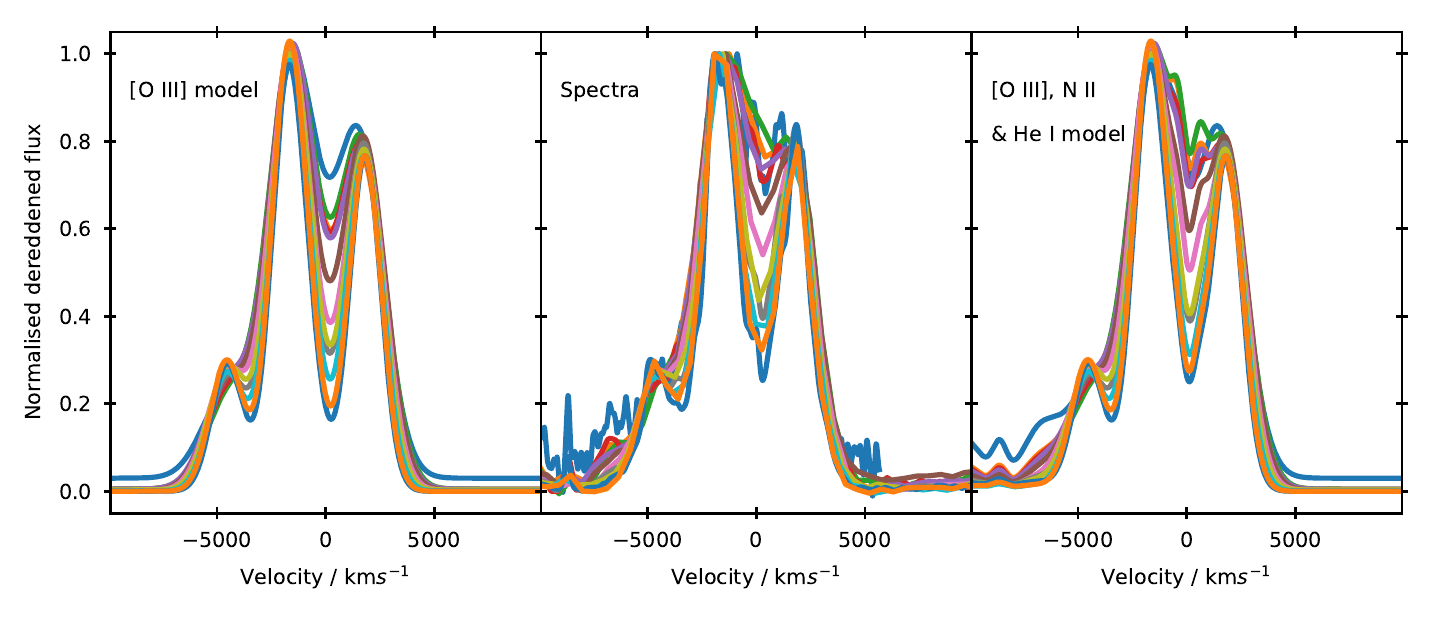}
\end{center}
\caption{Evolution of the [\ion{O}{III}] 4959+5007\,\AA\ profile from days 77--212 post-eruption. \textbf{Centre:} observed line profiles. \textbf{Left:} [\ion{O}{III}] line profile models. \textbf{Right:} [\ion{O}{III}] models including emission from \ion{N}{ii} 5001\,\AA, \ion{He}{i} 5016\,\AA, and H$\beta$ ($-8727$\,km\,s$^{-1}$).\label{fig:O5007prefrozenin}}
\end{figure*}

\begin{figure*}
\begin{center}
\includegraphics[width=0.9\columnwidth]{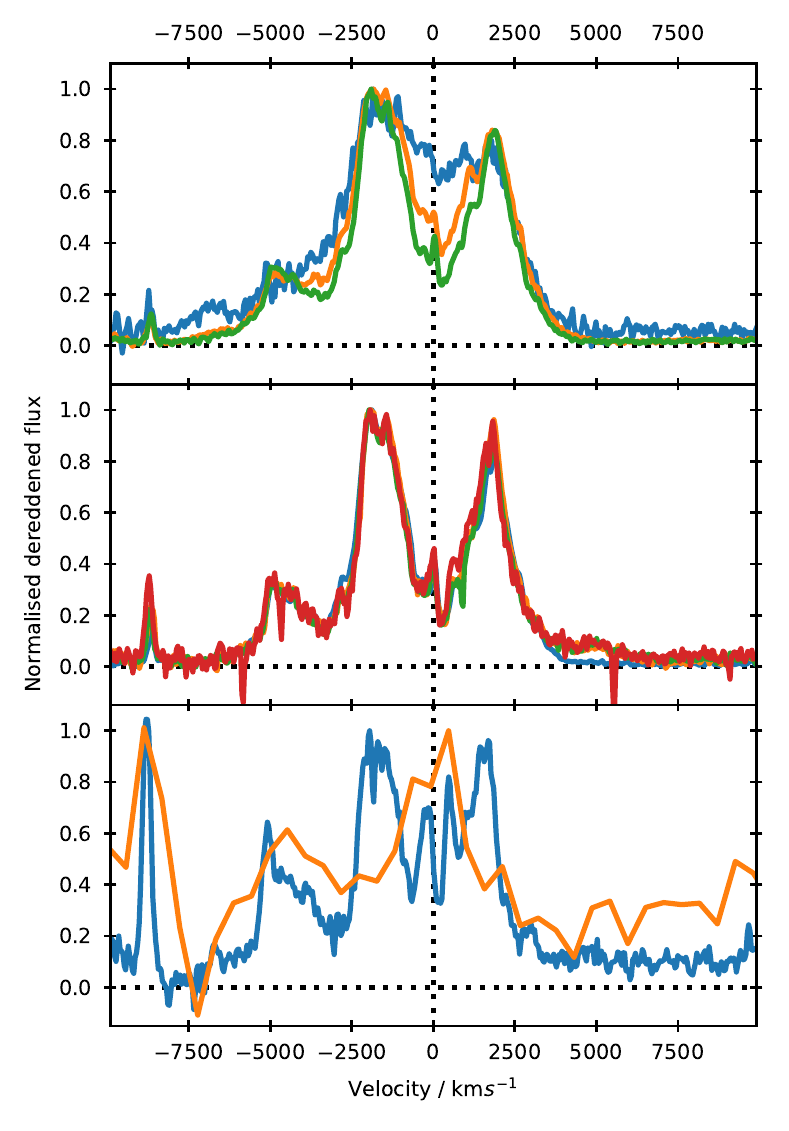}\hfill\includegraphics[width=0.9\columnwidth]{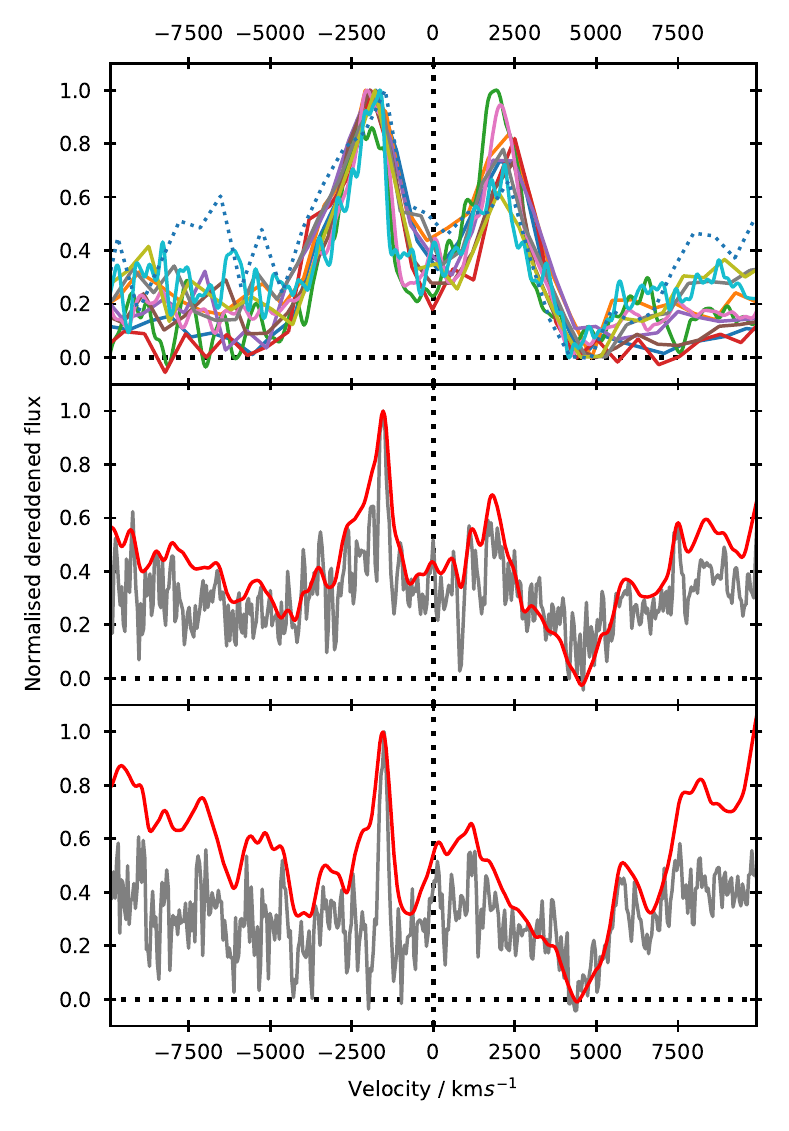}
\end{center}
    \caption{\textbf{Left:} Evolution of the high resolution [\ion{O}{iii}] 4959+5007\,\AA\ profile. Top panel, showing the transition to the 'frozen-in' state (as shown in low resolution in Figure~\ref{fig:O5007prefrozenin}). The middle panel shows profile `frozen-in' between days 220-351 post-eruption. The bottom panel shows time-averaged data for the late nebular phase (blue), where H$\beta$ ($-8727$\,km\,s$^{-1}$) is now similar in strength to [\ion{O}{iii}] 5007\,\AA. The orange line shows the post-nova data from day 854, here only \ion{He}{i} 5016\,\AA\ remains. \textbf{Right:} [\ion{O}{iii}] 4363\,\AA\ low resolution profile evolution, SPRAT data are shown at their native resolution, while Hiltner 2.4\,m data have been Gaussian smoothed to match. The top panel shows data from days 82--212, where the profile is frozen-in. The middle and bottom panels show the average profile between days 220--351 and 448--854, respectively for the low resolution (red) and high resolution (grey) spectra. By this time, H$\gamma$ has reasserted its dominance (at $\sim-1600$\,km\,s$^{-1}$).}
    \label{fig:O5007O4363Comp}
\end{figure*}

The combined [\ion{O}{iii}] 4959+5007\,\AA\ nebular emission complex was visible from the initial post-first Sun constraint spectrum, and dominant throughout the nebular phase (see Figures \ref{fig:Calspecmid} and \ref{fig:Calspeclate}). Although appearing to mirror the triple-peaked H$\alpha$ profile, the [\ion{O}{iii}] complex consists of a pair of overlapping double-peaked line profiles, the brighter centred at 5007\,\AA\ the fainter at 4959\,\AA\ (see Figures \ref{fig:O5007prefrozenin} and \ref{fig:O5007O4363Comp}). There is no evidence for a central component and, by this time, the central component seen in the permitted lines was only $\sim\!60$\,km\,s$^{-1}$ wide and therefore not associated with the ejecta. 

The nebular [\ion{O}{iii}] flux was measured by modelling the 5007\,\AA\ component with a symmetric pair of Gaussians offset equally either side of the rest wavelength. The 4959\,\AA\ was simultaneously modelled by scaling the 5007\,\AA\ profile. The blueward peak from the 4959\,\AA\ profile overlaps with the redward outer peak of H$\beta$. Thus, to de-blend [\ion{O}{iii}] and H$\beta$, we included H$\beta$ (based on the H$\alpha$ profile) in the nebular [\ion{O}{iii}] model (see Section \ref{sec:Balmerlines}). In addition, we incorporated \ion{N}{ii} 5001\,\AA\ and \ion{He}{i} 5016\,\AA\ lines using single Gaussians with widths matching the central H$\alpha$ peak.

Figure~\ref{fig:O5007prefrozenin} shows the evolution of the nebular [\ion{O}{iii}] 4959+5007\,\AA\ profile (days 77--212 post-eruption) as it transitions toward the `frozen-in' state. Initially, the space between the two 5007\,\AA\ peaks was `filled' by emission from \ion{N}{ii} 5001\,\AA\ and \ion{He}{i} 5016\,\AA. As the relative strength of [\ion{O}{iii}] increased the impact of \ion{N}{ii} and \ion{He}{i} diminished.

The left hand panel of Figure \ref{fig:O5007O4363Comp} presents the nebular [\ion{O}{iii}] profile in higher resolution. The top plot shows the pre-frozen evolution seen in Figure~\ref{fig:O5007prefrozenin}. The middle plot shows the frozen-in phase between days 220--351. The bottom plot (days 448--854) shows weakened [\ion{O}{iii}] (blue; with \ion{N}{ii}, \ion{He}{i}, and H$\beta$ again evident). Here, the orange line shows the low resolution profile obtained 854\,days post-eruption, when [\ion{O}{iii}] was no longer detectable.  

The flux evolution of [\ion{O}{iii}] 4959\,\AA\ and [\ion{O}{iii}] 5007\,\AA\ is shown in Figure~\ref{fig:Fluxes350d_newflux}. The first flux measurements are from day 77, yielding $(2.72\pm 0.17) \times 10^{-10}$\,erg\,s$^{-1}$\,cm$^{-2}$ and $(9.45\pm0.26) \times 10^{-10}$\,erg\,s$^{-1}$\,cm$^{-2}$ for 4959\,\AA\ and 5007\,\AA, respectively (a ratio of $3.5\pm0.2$). We fit the flux of both the 4959\,\AA\ and 5007\,\AA\ contributions by linking  both to a power law with index $-1.88\pm 0.10$. From day 346, the rate of decline steepened and the flux ratio between the components began to decrease.

\subsection{Auroral [O \textsc{iii}] 4363 \AA}

Similar to its nebular cousins, auroral [\ion{O}{iii}] 4363\,\AA\ was visible in the initial post-first Sun constraint spectra, and this emission rivalled the H$\alpha$ line. The line profile evolution is shown in the right-hand panels of Figure~\ref{fig:O5007O4363Comp}. The top plot presents days 82-212 and, unlike the nebular lines, the auroral profiles are already frozen-in. The signal-to-noise for the auroral profile rapidly diminished, as such, the middle and bottom panels simply show time averaged low- (red) and high-resolution (grey) spectra between days 220--351 and 448--854, respectively. While the low-resolution spectra suggest changes in the relative intensity of the two components, in the high-resolution spectra (days 220--351) we see that the auroral structure has faded with narrow H$\gamma$ emission becoming more dominant. By the post-nova phase, evidence for auroral [\ion{O}{iii}] has largely disappeared leaving just the narrow H$\gamma$ line.

The [\ion{O}{iii}] 4363\,\AA\ flux was measured by fitting the profile with a similar model to [\ion{O}{iii}] 5007\,\AA\ combined with a H$\alpha$-based profile for H$\gamma$. We also incorporated \ion{He}{i} 4388\,\AA\ emission using a single Gaussian matched to the width of the H$\alpha$ central peak. The flux measurements for [\ion{O}{iii}] 4363\,\AA\ are shown in Figure~\ref{fig:Fluxes350d_newflux}. The first measurement (day 77) yielded $(1.76\pm0.33) \times 10^{-10}$\,erg\,s$^{-1}$\,cm$^{-2}$, with the flux evolution described by a power law with index $-2.72\pm 0.46$.

\subsection{Other P-class neon novae}\label{sec:P class}

Four P-class novae listed in \cite{Strope2010} showed neon lines in their spectra, and their $t_2$ decline times ranged from 1\,day (V838\,Her) to 20\,days (QU\,Vul). \ion{Fe}{ii} emission was present in the early spectra of QU\,Vul \citep{Rosino1992} and V1974\,Cyg \citep{Chochol1993,Rafanelli1995}. There was no mention of \ion{Fe}{ii} in the spectra of V838\,Her \citep{Vanlandingham1996}, and spectra of V351\,Pup were only available from 136 days after eruption \citep{Saizar1996}. Expansion velocities decreased with decline time, and ranged from $1700-4200$\,km\,s$^{-1}$, with double P\,Cygni profiles evident in the Balmer lines in early spectra of QU\,Vul, V1974\,Cyg and V838\,Her \citep{Chochol1993,Rafanelli1995,Rosino1992,Vanlandingham1996}. V1974\,Cyg is a proposed magnetic CV, with $P=0.08$\,d \citep{Chochol1997}. During the early evolution, there are no substantial differences between the spectra of these four and V392\,Per. However, there is no evidence for similar late-time, narrow-line, behaviour in these systems (see Section~\ref{sec:xray_and_accretion}).

\subsection{X-ray spectral modelling}\label{sec:Xrayfit}

\begin{figure*}
\begin{center}
\includegraphics[width=0.8\textwidth]{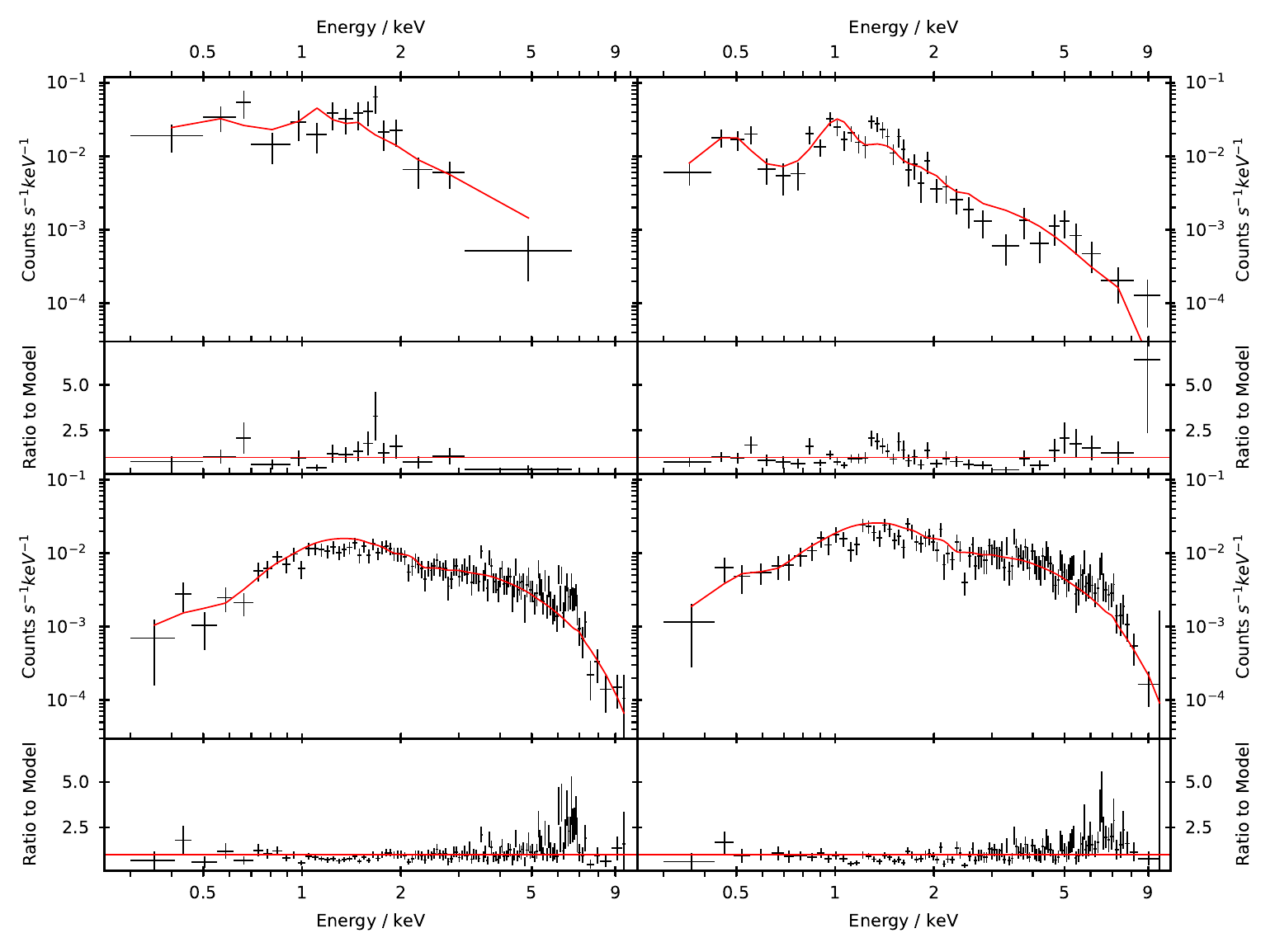}
\end{center}
\caption{{\it Swift}/XRT spectra of V392\,Per (black) and best-fit models (red), residuals included underneath. {\bf Top left:} single observation, 83\,days post-eruption. {\bf Top right:} combination of observations 89--97\,days post-eruption. {\bf Bottom left:} days 112--361 days post-eruption. {\bf Bottom right:} days 449--849 post-eruption.}
\label{fig:Spec_comb}
\end{figure*}

Figure \ref{fig:Spec_comb} shows {\it Swift}/XRT spectra from four epochs. For each epoch the upper panel shows the data (black) and best fit model (red), while the lower panel shows model residuals as a ratio. The XRT spectra were fitted using a combination of a black-body and collisionally excited plasma (APEC) components, where appropriate. All spectra were modelled using a fixed column \citep[$N_\mathrm{H}=4.8\times 10^{21}\,\mathrm{cm}^{-2}$; equivalent to $E(B-V)=0.7$, converted using Equation 1 from][]{2009MNRAS.400.2050G}. Table~\ref{tab:SpectraNH0.6myerrors} summarises the results of the fitting. We note that if $N_\mathrm{H}$ were permitted to vary freely, larger values ($\sim10^{22}\,\mathrm{cm}^{-2}$) were obtained (see Section~\ref{sec:xray_and_accretion}). The choice of column makes little difference to the resulting findings.

\begin{table*}
\caption{Fits to \textit{Swift}/XRT spectra with $N_\mathrm{H}=4.8 \times 10^{21}\,\mathrm{cm}^{-2}$. The third column shows the required model components. Subsequent columns record the temperature and normalisation of the required components, black-body, first APEC, and second APEC, respectively. The final column reports the goodness of fit, the (modified) Cash statistic per degree of freedom (C stat / d.o.f.).} 
\label{tab:SpectraNH0.6myerrors}
\begin{tabular}{llllllllll}
\hline
Time & Exp.\ time & Components & $kT_\mathrm{BB}$ & Norm$_\mathrm{BB}$  & $kT_\mathrm{APEC1}$ & Norm$_\mathrm{APEC1}$  & $kT_\mathrm{APEC2}$ & Norm$_\mathrm{APEC2}$  & C stat / d.o.f. \\
/ days & / ks & & / eV & / $10^{-3}$ & / keV & / $10^{-3}$ & / keV & / $10^{-3}$ \\
\hline

\phantom{0}83 & \phantom{0}1.6 & bb+apec & $62^{+17}_{-14}$ & $0.8^{+3.0}_{-0.6}$ & $2.3^{+1.2}_{-0.5}$ & $2.0^{+0.4}_{-0.4}$ & \ldots & \ldots & 87 / 74 \\

\phantom{0}89--97 & 12.0 & bb+apec+apec & $48^{+10}_{-8}$ & $2^{+5}_{-1}$ & $1.2^{+0.2}_{-0.1}$ & $0.5^{+0.2}_{-0.1}$ & $>\phantom{0}4.2$ & $0.5^{+0.2}_{-0.1}$ & 208 / 192 \\

112--361 & 43.6 & apec+apec &  \ldots & \ldots & $<0.06$ & $<2000$ & $>58.3$ & $1.93^{+0.08}_{-0.08}$ & 667 / 587 \\

449--849 & 19.8 & apec+apec &  \ldots & \ldots & $0.08^{+0.08}_{-0.05}$ & $14^{+757}_{-14}$ & $>57.2$ & $3.0^{+0.1}_{-0.1}$ & 548 / 537 \\ 
\hline
\end{tabular} 
\end{table*}

The spectra at the top of Figure~\ref{fig:Spec_comb} show the XRT data taken 83 days post-eruption (left) and the combined spectra between days 89--97 (right), both taken after exiting from the first Sun constraint. On day 83, the count rate was $0.06\,\mathrm{s}^{-1}$ and the spectrum was relatively soft, with a HR of 2.7. By days 89--97 the count rate had dropped to $0.035\,\pm\,0.003\,\mathrm{s}^{-1}$, and the HR had hardened to $4.4$ and then to $6.3$. These spectra correspond to the first three data points in Figure~\ref{fig:SwiftXRTCounts_new} and are the softest X-ray spectra taken of V392\,Per. These first two spectra show clear count rate excesses at low energies and, as such, we fitted the earliest spectrum using the combination of a black-body ($k_\mathrm{B}T=62^{+17}_{-14}$\,eV) and hot plasma, APEC, component. The day 89--97 spectra required an additional and hotter APEC component to account for the emission $>5$\,keV that is not seen in the day 83 spectrum, likely due to the shorter integration. Given these spectra, the declining soft X-ray flux during this period, and the optical spectra at the time, we conclude that we are observing the tail of the super-soft source stage of V392\,Per, and that the SSS ended at $97\,\mathrm{d}<t_\mathrm{SSS,off}<112\,\mathrm{d}$. The spectra at the bottom of Figure~\ref{fig:Spec_comb} show the XRT data between days 112--361 (left) and 449--849 (right). These are harder and clearly lack the SSS component. Here, both spectra are modelled using a pair of APEC components, with the best fit temperatures unchanged between the two epochs, although the count rate is higher at later times.

\section{Discussion}\label{sec:Disc}

V392\,Per is the first pre-known DN to be observed as a $\gamma$-ray bright CN. Here, we aggregate our reported observations to present a plausible description of the underlying system. 

\subsection{A shock-powered light curve?}\label{sec:shock_lc}

\begin{figure}
\begin{center}
	\includegraphics[width=0.9\columnwidth]{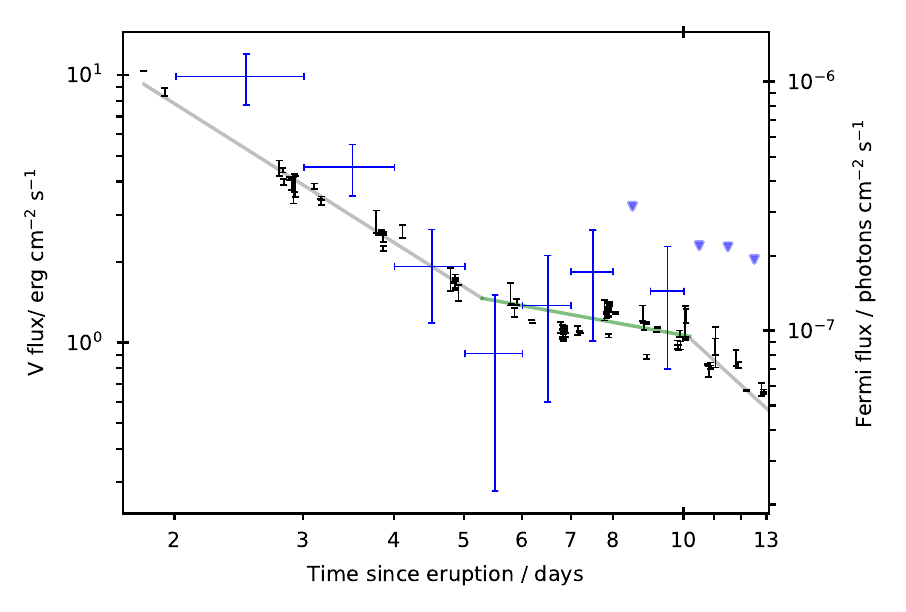}
	\end{center}
    \caption{$V$-band light curve (in black), overlaid with Fermi-LAT $\gamma$-ray light curve (in blue).} 
    \label{fig:VLC_Fermi}
\end{figure}

The Fermi-LAT detection of $\gamma$-ray emission from V392\,Per \citep{2022arXiv220110644A} followed soon after the optical detection. However, $\gamma$-ray emission might have been detected earlier were it not for technical problems with Fermi. In Figure~\ref{fig:VLC_Fermi} we directly compare the Fermi-LAT flux \citep[from][]{2022arXiv220110644A} with the $V$-band flux (see Section~\ref{sec:LCfitting}). As has been reported for other $\gamma$-ray novae \citep{Ackermann14,Aydi2020}, there appears to be a clear correlation between the $\gamma$-ray and optical emission during the early evolution.

The early $\gamma$-ray and optical declines both follow similar power-laws until day $\sim5$, when both fluxes plateau. As the optical plateau ends (day $\sim10$) and the decline resumes, the Fermi-LAT detections cease (although observations continued). Here, we propose that the early (pre-first Sun constraint) optical light curve is driven by the evolution of shocks between and within multiple mass ejection components \citep[as discussed in][]{Gordon2021}:

As reported in Section~\ref{sec:Pcygni}, there is evidence for multiple mass ejections, an initial event with $v\sim3000$\,km\,s$^{-1}$ being swept up and shocked 2.5\,d post-eruption by a faster ejecta at $v\sim5000$\,km\,s$^{-1}$ --- corresponding with the initial Fermi-LAT detection. These shocks will have accelerated ions to relativistic velocities, which emitted $\gamma$-rays while interacting with particles or photons in the surroundings \citep{Martin2018,Aydi2020}. While $\gamma$-ray emission from novae can also be linked to the ejecta shocking and sweeping up pre-existing circumbinary material \citep[e.g., a red giant wind;][]{Cheung2014}, here we see no evidence for the associated coronal emission lines \citep{1987rorn.conf...27R} or sustained bulk ejecta deceleration expected in such cases \citep{1985MNRAS.217..205B,2016ApJ...833..149D}. As such, we propose that the most likely source of the initial $\gamma$-ray emission is inter-ejecta shocks between these two components. With the lack of very early optical data, evidence for an additional, earlier, light curve peak corresponding to the initial ejection is unavailable.

The spectral evolution during the initial plateau is complex, and is additionally challenging due to the decreasing optical depth likely to be simultaneously occurring. The light curves of many novae enter quasi-plateau phases around $t_3$. For the recurrents, this has been proposed to be driven by a surviving, or rapidly reformed, accretion disk emerging from the receding photosphere, with the unveiling of the SSS occurring toward the end of the plateau as the inner disk is revealed. However, unlike recurrent nova plateaus \citep[e.g.,][]{2018ApJ...857...68H}, here we see no evidence for \ion{He}{ii} emission during the plateau -- which would be expected from a disk. Indeed, \ion{He}{ii} emission is only seen after the first Sun constraint during the nebular phase. The plateau corresponds with the end of the clear `bulk' ejecta merger, but there remains evidence for on-going `minor' interaction. We tentatively propose that the $\gamma$-ray emission during the plateau is driven by intra-ejecta shocks following the major merger event. 

\subsection{X-ray emission and accretion}\label{sec:xray_and_accretion}

In Section~\ref{sec:Xrayfit} we presented evidence that the end of the SSS X-ray emission phase was caught just as V392\,Per emerged from its first Sun-constraint. The relationships in \citet{2014A&A...563A...2H} predict a SSS turn-on time $t_\mathrm{SSS,on}=13^{+6}_{-4}$\,days (based on $t_2$), a corresponding turn-off $t_\mathrm{SSS,off}=60^{+80}_{-40}$\,days, and a SSS black-body parameterised temperature of $k_\mathrm{B}T\sim90$\,eV. A caveat here is that the \citet{2014A&A...563A...2H} relationships are defined for the M\,31 nova population, where deep X-ray observations are prohibitive due to distance, and as such they may systematically over- and under-predict $t_\mathrm{SSS,on}$ and $t_\mathrm{SSS,off}$, respectively, for Galactic novae. Nonetheless, these predictions are compatible with the available X-ray observations of V392\,Per. We again note that there was no associated optical spectral evidence of a SSS before the first Sun-constraint.

As is demonstrated in Figure~\ref{fig:Spec_comb}, at all observed times, there is a substantial contribution to the X-ray luminosity by a harder component, that is well described by a single, or pair of, hot collisional plasma (APEC) models. Such emission is often associated with shocks. However, the consistency and longevity of the hard emission -- from at least day 83 to beyond day 800 -- reveals that this emission cannot be associated with an expanding ejecta. We do note that, unfortunately, there is no pre-eruption X-ray or UV data available for comparison. 

During the post-nova phase, we see clear, strong, and very narrow emission from \ion{H}{i}, \ion{He}{i}, and especially \ion{He}{ii}, on top of a blue continuum -- indicative of an accretion disk. However, the $<100$\,km\,s$^{-1}$ width of these lines might imply a disk very close to face-on and would seemingly contradict the orbital modulation observed (see Section~\ref{sec:orbper}), which suggests an inclination closer to edge-on. At the same time, the SED of V392\,Per (see Figure~\ref{fig:SEDaverage}) shows strong and consistent emission in the near-UV. As shown in Figure~\ref{fig:SwiftXRTCounts_new}, the near-UV and X-ray emission appear correlated. We conclude from this that the UV and X-rays arise from the same system component and, given the shape of the SED, this will be from a reformed accretion disk. Most quiescent CNe do not show substantial X-ray emission, suggesting that V392 Per is a magnetic CV. 

However, the potential orbital period ($P\simeq3.2$\,d; Section \ref{sec:orbper}) is too long to be consistent with a tidally-locked polar configuration \citep{Mukai2017}. Therefore, with APEC temperatures, perhaps, in excess of 50\,keV, the X-ray emission is similar to that expected to emanate from the standing/standoff shocks observed in the accretion environment surrounding intermediate polars (IP); CVs with WD magnetic fields in the range $10^6 \lesssim B \lesssim 10^7$\,G. A field of this strength would truncate the inner part of the accretion disk, causing the accreted material to flow along the magnetic field lines in accretion curtains onto the WD. Could the presence of an accretion curtain be behind the higher than expected column seen in the X-ray fits (see Section~\ref{sec:Xrayfit})? We note that neither our $i'$-band high cadence data, or \textit{Swift} data, are suitable for searching for a signal from the WD spin period -- a key diagnostic of an IP.

We have linked the narrow emission line spectrum in the post-nova phase to active accretion in the system. As such, we restate that this disk spectrum was already visible once the system had exited the first Sun-constraint (e.g. Figure~\ref{fig:HeIIprofiles}), while the nova ejecta was still fading, and while the SSS was still on. As such, the disk must have (at least) partially survived the eruption, or reformed during the SSS phase. From this we infer that there could have been active accretion during the SSS-phase, potentially `re-fuelling' the WD and prolonging the SSS phase \citep[cf.][]{2018MNRAS.474.2679A,2018ApJ...857...68H}.

\subsection{Pre-nova versus post-nova}

In Figure~\ref{fig:post_vs_pre_nova}, we directly compared the `steady state' post-nova luminosity with the pre-nova AAVSO light curve. The pre-nova state shows a quiescent baseline at $V\sim17$\,mag (low-state), punctuated with several 2--3\,mag amplitude DN outbursts (high-state). The timescale of these DN outbursts are more akin to those seen in longer orbital period systems, such as GK\,Per. As also discussed by \citet{Munari2020}, V392\,Per has not returned to its pre-nova quiescent level. The system has remained at an elevated high-state of $V\sim15$\,mag for (at least) two years post-eruption. During this time, the near-UV and X-ray luminosity continue to creep upwards (Figure~\ref{fig:XRT_uvw2_new}) -- i.e., there is no evidence so far that the system will return to the low-state. In DN systems, the majority of accretion onto the WD surface occurs when the disk is in a high state, i.e., during a DN outburst. Here, we infer that V392\,Per is maintaining an elevated level of accretion post-nova, and that it is currently best classified as a nova-like variable, rather than a DN. It is unclear as to why V392\,Per is remaining in this post-nova high-state. It may simply be that irradiation of the donor by the recent nova eruption is driving elevated mass loss from the companion \citep[cf.\ T\,Pyx;][]{2021MNRAS.507..475G}. We are not currently in a position to predict when, or even if, the system will revert to its pre-nova state.

\subsection{The underlying system}\label{sec:underlying_system}

With a $t_2$ as short as 2\,days and a SSS turn-off of $\sim100$\,days, the indications are that the WD in this system is particularly massive \citep[see, e.g.,][]{Schwarz2011}. Indeed, the V392\,Per nova eruption is one of the fastest evolving on record, and the SSS phase may have been unusually extended through refuelling by a surviving or rapidly reformed disk. Novae with similar parameters are expected to host WDs with masses in excess of 1.1\,M$_\odot$, perhaps up to 1.3\,M$_\odot$ \citep[see][and references therein]{2005ApJ...623..398Y,2011ApJ...727..124O,Hillman2016}. The relatively high SSS black-body temperature ($k_\mathrm{B}T\sim50$\,eV) seen even at the very end of the SSS phase is similarly suggestive of a massive WD. With strong forbidden Ne lines present in the spectra \citep[see Section~\ref{sec:specresults} and][]{Munari2020ATel} there is a distinct possibility that V392\,Per hosts a massive ONe WD.

\begin{figure}
\begin{center}
\includegraphics[width=0.9\columnwidth]{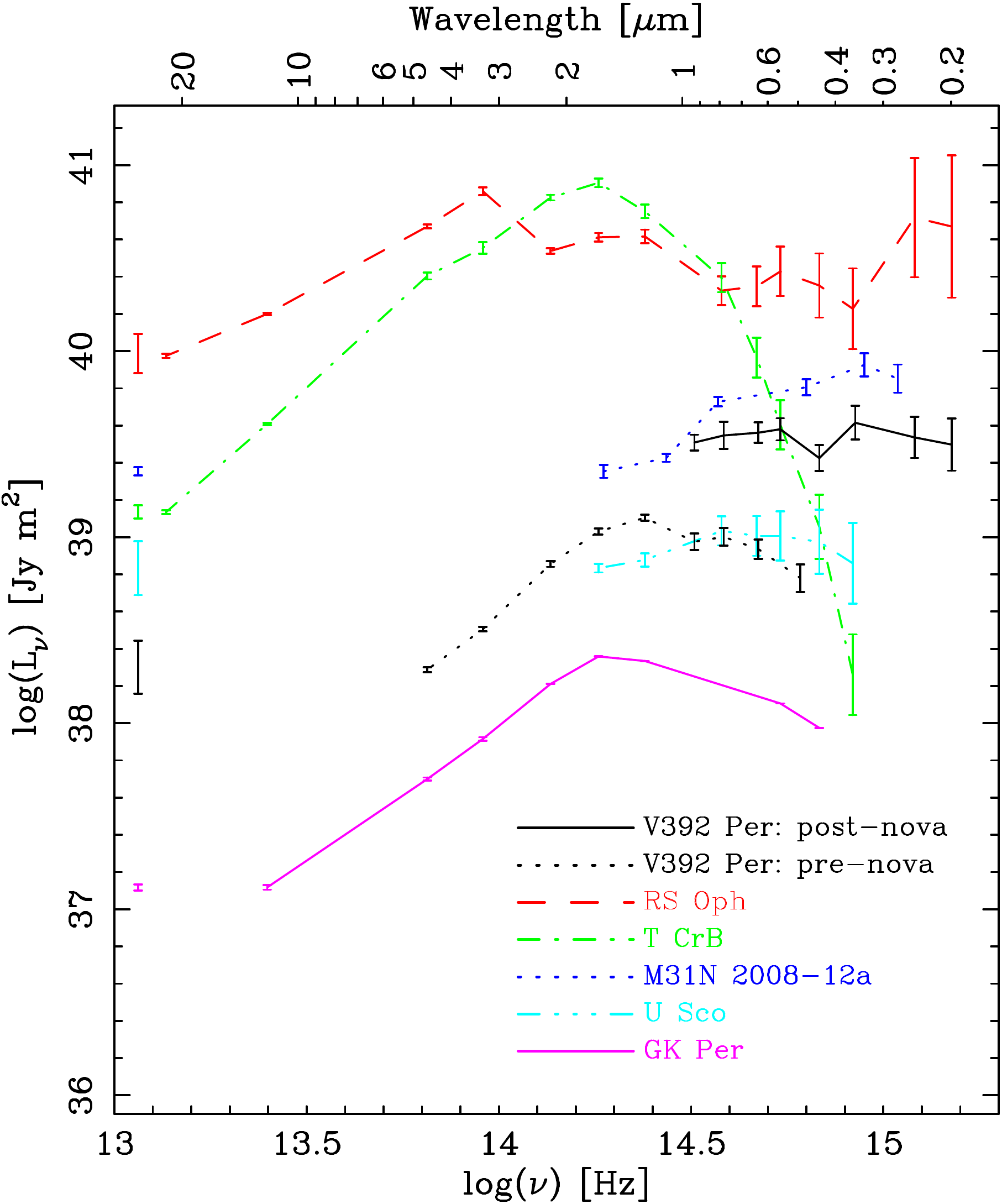}
\end{center}
\caption{Distance and extinction corrected quiescent SEDs of V392\,Per, RS\,Oph, T\,CrB, M31N\,2008-12a, U\,Sco, and GK\,Per. Error bars include photometric and extinction uncertainties; distance uncertainties indicated to the left of the plot, lines are to aid the reader. Data from this work, \citet{Darnley2012,2017ApJ...849...96D}, \citet{2014MNRAS.444.1683E}, \citet{2021A&A...649A...1G}, \citet{Munari2020}, \citet{2006AJ....131.1163S}, and \citet{Page2022}.\label{fig:SED_cmp}}
\end{figure}

In Figure~\ref{fig:SED_cmp} we present an updated V392\,Per SED \citep[cf.][]{DarStar18}. The dotted lower-luminosity black line shows the pre-nova SED, ranging from the Wise B2 band up to the $g$-band \citep[data from][and references therein]{DarStar18,Munari2020}. As extensively discussed by \citet{Munari2020}, these pre-nova data suggest a warm and cool component. \citeauthor{Munari2020} utilised these data to constrain the donor, arriving at a similar (albeit more detailed) conclusion to \citeauthor{DarStar18} that the donor is likely a sub giant or low luminosity giant \citep[specifically:\ G9\,IV-III; 5.34\,R$_\odot$; 1.35\,M$_\odot$; 15\,L$_\odot$; $T_\mathrm{eff}=4875$\,K;][]{Munari2020}. With the addition of the Wise mid-IR data, we find that the pre-nova data are reasonably well represented by a single black-body, with $T_\mathrm{eff}=5700\pm400$ \citep[cf.\ $\simeq5000$\,K for GK\,Per;][]{2021MNRAS.507.5805A}, $R=7.8\pm0.6$\,R$_\odot$, and $L=55^{+20}_{-18}\,\mathrm{L}_\odot$. 

The pre-nova SED is remarkably similar to that of GK\,Per (magenta data), albeit $\sim10$ times as luminous. It is easy to draw comparisons between the two systems: both novae have evolved companions and long orbital periods; DN outbursts characterised by their month-long longevity; and like GK\,Per, V392\,Per may be an IP. However, their post-nova behaviour is very different:

The post-nova SED is indicated by the solid black line in Figure~\ref{fig:SED_cmp}, here the data span the $z'$-band to the \textit{Swift}/UVOT uvw2 near-UV filter. It is clear that the post-nova emission is substantially greater than that seen pre-eruption (in the low states, at least). Here, the post-nova SED is reminiscent of, and indeed of similar luminosity to, the disk in M31N\,2008-12a \citep{2017ApJ...849...96D}. This implies that $\dot{M}$ during the post-nova phase may be very high, which may act to lessen the time toward the next eruption.

\section{Summary and Conclusions}\label{sec:Conc}

V392\,Per is a known CV, which exhibited month-long GK\,Per-like DN outbursts, and its only known classical nova eruption was discovered on 2018 April 29. Panchromatic photometric and spectroscopic follow-up took place, with optical observations intensifying after the reported detection of $\gamma$-rays by Fermi-LAT, although the system was already in \textit{Swift} Sun constraint at eruption. Post-Sun constraint, the eruption had entered the nebular spectral phase and \textit{Swift} observations began. Since $\sim250$\,days post-eruption, V392\,Per has remained in a high-state, consistently $\sim2$\,mag brighter than the pre-eruption quiescent minimum. Here we summarise our key findings:

\begin{enumerate}
\item \textit{Gaia} EDR3 astrometry indicates $d=3.5^{+0.6}_{-0.5}$\,kpc, and we derive $E\left(B-V\right)=0.70^{+0.03}_{-0.02}$.
\item With $t_2=2.0\pm0.2$\,days, the eruption is classed as `very fast', indicative of a high mass WD.
\item The early spectra indicate that V392\,Per is a rare \ion{Fe}{ii}-broad class, with ejection velocities up to $5000$\,km\,s$^{-1}$.
\item Evolution of early-time \ion{H}{i} P\,Cygni profiles strongly suggest there were two distinct mass ejections, with the higher velocity second ejecta running into and shocking the first.
\item These inter-ejecta, and subsequent intra-ejecta, shocks drove the $\gamma$-ray emission.
\item Distinct similarities between the $\gamma$-ray and early-optical evolution suggest that the early luminosity was powered by the shock emission. 
\item The X-ray observations indicate the SSS was turning off as V392\,Per emerged from Sun constraint on day 83.
\item Inferred SSS parameters along with forbidden Ne lines also suggest a high mass, perhaps ONe, WD.
\item Optical spectra show two distinct contributions: a broad initially triple, then double peaked fading ejecta spectrum; and a narrow lined and persistent accretion disk spectrum.
\item Persistent hard X-ray emission, and post-nova near-UV luminosity, is consistent with continuing accretion, suggesting that V392\,Per is an intermediate polar.
\item Post-nova high cadence $i'$-band data indicate an orbital period of $P=3.230\pm0.003$\,days.
\item The pre-nova mid-IR--optical SED suggests a sub-giant or low luminosity giant donor.
\item The post-nova optical--NUV SED is substantially more luminous and is akin to an accretion disk.
\end{enumerate}

In a follow-up work, we will use the extensive spectra published here to explore the underlying geometry and ionisation structure of the V392\,Per ejecta. We leave V392\,Per in a post-nova elevated high-state, with no indication that the system will return to its pre-nova low-state and DN behaviour. Our observations continue\ldots

\section*{Acknowledgements}

We would like to express our gratitude to Dr Mike Shara for his helpful and thoughtful comments when refereeing the original manuscript. FMG acknowledges a PhD studentship from the Science and Technology Funding Council (STFC). MJD and EJH receive funding from STFC. KLP acknowledges funding from the UK Space Agency. The Liverpool Telescope is operated on the island of La Palma by Liverpool John Moores University in the Spanish Observatorio del Roque de los Muchachos of the Instituto de Astrofisica de Canarias with financial support from STFC. This work uses observations from the Las Cumbres Observatory global telescope network. This work uses observations from the MDM Observatory, operated by Dartmouth College, Columbia University, Ohio State University, Ohio University, and the University of Michigan. We would like to thank Justin Rupert for obtaining many of the MDM spectra during regularly scheduled monthly service blocks. The LBT is an international collaboration among institutions in the United States, Italy and Germany. LBT Corporation partners are:\ The University of Arizona on behalf of the Arizona university system; Istituto Nazionale di Astrofisica, Italy; LBT Beteiligungsgesellschaft, Germany, representing the Max-Planck Society, the Astrophysical Institute Potsdam, and Heidelberg University; The Ohio State University; and the collaborating institutions The University of Notre Dame, University of Minnesota, and University of Virginia. This work has made use of data from the European Space Agency (ESA) mission {\it Gaia} (\url{https://www.cosmos.esa.int/gaia}), processed by the {\it Gaia} Data Processing and Analysis Consortium (DPAC, \url{https://www.cosmos.esa.int/web/gaia/dpac/consortium}). Funding for the DPAC has been provided by national institutions, in particular the institutions participating in the {\it Gaia} Multilateral Agreement. We acknowledge with thanks the variable star observations from the AAVSO International Database contributed by observers worldwide and used in this research. We thank Dr Mike Healy and Prof.\ Phil James for feedback on an early version of this manuscript, Dr Andy Beardmore for discussion on magnetic CVs, and Prof.\ Brad Schaefer for discussion regarding TESS. We thank Ian Miller and Erik Schwendeman, sadly deceased, for their observations of V392\,Per, submitted to the AAVSO database.

\section*{Data Availability}

This work was funded by UKRI grants ST/S505559/1 and ST/V00087X/1. The Liverpool Telescope was funded by UKRI grants ST/S006176/1 and ST/T00147X/1. For the purpose of open access, the author has applied a Creative Commons Attribution (CC BY) licence to any Author Accepted Manuscript version arising. All data supporting this study are available from the respective telescope archives, in the supplementary information, or on reasonable request from the author. 

\bibliographystyle{mnras}
\bibliography{V392Pernewbib} 

\clearpage
\onecolumn 
\appendix

\section{Supplementary on-line only material}

\begin{longtable}[]{lllll}
\caption{Log of spectral observations of V392\,Per. For FRODOSpec and LBT/MODS observations, the resolution of the red arm and the blue arm are both given. }\label{tab:Specobs} \\
\hline
$\Delta t$  & Date (UT) & Telescope/ & Resolution & Exp. \\ 
(d) & & instrument & $\sim$R & time (s) \\
\hline
\endfirsthead

\multicolumn{5}{c}{\tablename\ \thetable{} -- continued from previous page} \\
\hline
$\Delta t$  & Date (UT) & Telescope/ & Resolution & Exp. \\ 
(d) & & instrument & $\sim$R & time (s) \\
\hline
\endhead

\hline
\multicolumn{5}{r}{Continued on next page}
\endfoot

\endlastfoot

1.9 & 2018-04-29.898 & RLE /THO-UK & 527 & 703 \\
2.1 & 2018-04-30.116 & Hiltner/OSMOS & 1600 & 7$\times$40 \\
2.9 & 2018-04-30.868 & RLE /THO-UK & 530 & 636 \\
2.9 & 2018-04-30.889 & RLE /THO-UK & 530 & 636 \\
3.8 & 2018-05-01.836 & EBE/SSO-FR & 503 & 1235 \\
3.9 & 2018-05-01.863 & DDJ/LSS-FR & 525 & 770 \\
4.2 & 2018-05-02.154 & JPE/DHO-US & ... & 838 \\
4.8 & 2018-05-02.794 & BER/BVO-IT & 3975 & 484 \\
\end{longtable}

\clearpage

\begin{longtable}[]{llllllll}
\caption{The Pan-STARRS reference stars used to calibrate the photometry of V392\,Per. The \textit{Swift}/UVOT $U$ magnitude of \#15 is included. Errors on magnitudes for $BVr'i'z'$ are $\pm 0.034$, $\pm 0.012$, $\pm 0.004$, $\pm 0.005$ and $\pm 0.010$, respectively. }\label{tab:refphot} \\
\hline
Reference  & Pan-STARRS & $U$ & $B$ & $V$ & $r'$ & $i'$ & $z'$  \\
Star no & Object ID & / mag & / mag & / mag & / mag & / mag & / mag \\
\hline
\endfirsthead

\multicolumn{8}{c}{\tablename\ \thetable{} -- continued from previous page} \\
\hline
Reference  & Pan-STARRS & $U$ & $B$ & $V$ & $r'$ & $i'$ & $z'$  \\
Star no & Object ID & / mag & / mag & / mag & / mag & / mag & / mag \\
\hline
\endhead

\hline
\multicolumn{8}{r}{Continued on next page}
\endfoot

\hline
\endlastfoot
1   &   164840708125355000   &   \ldots   &   18.95   &   17.72   &   17.29   &   16.82   &   16.51 \\
2   &   164800708138869000   &   \ldots   &   20.16   &   18.96   &   18.54   &   18.08   &   17.70 \\
3   &   164810708164722000   &   \ldots   &   20.02   &   18.60   &   18.08   &   17.47   &   17.03 \\
4   &   164830708184131000   &   \ldots   &   19.38   &   18.07   &   17.60   &   17.03   &   16.64 \\
5   &   164820708189789000   &   \ldots   &   21.72   &   20.22   &   19.68   &   19.02   &   18.56 \\
6   &   164830708206900000   &   \ldots   &   22.53   &   21.01   &   20.46   &   19.77   &   19.23 \\
7   &   164830708274029000   &   \ldots   &   19.59   &   17.98   &   17.38   &   16.71   &   16.24 \\
8   &   164810708286245000   &   \ldots   &   19.29   &   18.00   &   17.54   &   17.01   &   16.63 \\
\end{longtable}

\clearpage

\begin{longtable}[]{llllll}
\caption{$u'$BV$r'i'z'$ and \textit{Swift}/UVOT photometry of V392\,Per (AB magnitudes).\label{tab:Bphot}} \\

\hline
Date (UT) & $\Delta$ t & Telescope  & Filter & Exposure time  & Photometry \\
 & / days &  \& Instrument &  & /s & /mag     \\
\hline
\endfirsthead

\multicolumn{6}{c}{\tablename\ \thetable{} -- continued from previous page} \\
\hline
Date (UT) & $\Delta$ t & Telescope  & Filter & Exposure time  & Photometry \\
 & / days &  \& Instrument &  & /s & /mag     \\
\hline
\endhead

\hline
\multicolumn{6}{r}{Continued on next page}
\endfoot

\hline
\endlastfoot

2018 Jul 21.192 & 84.202 & LT IO:O & $u'$ & $3\times 30$ & 14.751 $\pm$ 0.011 \\
2018 Jul 22.217 & 85.227 & LT IO:O & $u'$ & $3\times 30$ & 14.838 $\pm$ 0.008 \\
2018 Jul 23.194 & 86.204 & LT IO:O & $u'$ & $3\times 30$ & 14.900 $\pm$ 0.009 \\
2018 Jul 24.185 & 87.195 & LT IO:O & $u'$ & $3\times 30$ & 14.865 $\pm$ 0.012 \\
2018 Jul 25.182 & 88.192 & LT IO:O & $u'$ & $3\times 30$ & 14.973 $\pm$ 0.013 \\
2018 Jul 26.190 & 89.200 & LT IO:O & $u'$ & $3\times 30$ & 14.898 $\pm$ 0.015 \\
2018 Jul 27.182 & 90.192 & LT IO:O & $u'$ & $3\times 30$ & 14.832 $\pm$ 0.013 \\
2018 Jul 28.175 & 91.185 & LT IO:O & $u'$ & $3\times 30$ & 15.122 $\pm$ 0.012 \\
\end{longtable}

\clearpage

\begin{longtable}[]{llll}
\caption{AAVSO observers, including their country and astronomical society if available.}
\label{tab:AAVSOobservers} \\
\hline
Observer code  & Name & Country & Society\\
\hline
\endfirsthead

\multicolumn{4}{c}{\tablename\ \thetable{} -- continued from previous page} \\
\hline
Observer code  & Name & Country & Society\\
\hline
\endhead

\hline
\multicolumn{4}{r}{Continued on next page}
\endfoot

\hline
\endlastfoot
AAVSO AFSA    & Sold\'an Alfaro, Francisco  & ES      & AAVSO   \\
AAVSO ATE     & Arranz, Teofilo          & ES      & AAVSO   \\
AAVSO BDG     & Boyd, David              & GB      & BAA-VSS \\
AAVSO BMSA    & Bundas, Matthew          & US      & AAVSO   \\
AAVSO BRIA    & Biernikowicz, Richard    & PL      & AAVSO   \\
AAVSO CDJA    & Coulter, Daniel          & US      & AAVSO   \\
AAVSO DKS     & Dvorak, Shawn            & US      & AAVSO   \\
AAVSO DPA     & Diepvens, Alfons         & BE      & VVS     \\
\end{longtable}

\clearpage

\begin{table*}
\caption{V392\,Per light curve parameters, under the assumption that each can be modelled by up to six broken power laws of form $f\propto t^\alpha$, where $t_\mathrm{i}$ and $t_\mathrm{f}$ denote the initial and final extent of each power law, respectively, and $D$ is the duration of each power law's dominance.\label{tab:Fits_new}}
\begin{tabular}{lllll|lllll}
\hline
uvw2 & $\alpha$ & $t_\mathrm{i}$ [d] & $t_\mathrm{f}$ [d] & D [d] & uvm2 & $\alpha$ & $t_\mathrm{i}$ [d] & $t_\mathrm{f}$ [d] & D [d] \\ \hline
5 & -1.66 $\pm$ 0.14 & 70.0 & 186.4 & 116.4 & 5 & -1.64 $\pm$ 0.11 & 70.0 & 181.0 & 111.0 \\
6 & 0  & 186.4 & 849.4 & 663.1 & 6 & 0  & 181.0 & 849.4 & 668.4 \\ \hline
uvw1 & $\alpha$ & $t_\mathrm{i}$ [d] & $t_\mathrm{f}$ [d] & D [d] & u' fit & $\alpha$ & $t_\mathrm{i}$ [d] & $t_\mathrm{f}$ [d] & D [d] \\ \hline
4 & \ldots & - & \ldots & - & 4 & -1.93 $\pm$ 0.42 & 70.0 & 103.5 & 33.5 \\
5 & -1.82 $\pm$ 0.09 & 70.0 & 182.2 & 112.2 & 5 & -1.60 $\pm$ 0.10 & 103.5 & 195.9 & 92.3 \\
6 & 0  & 182.2 & 849.4 & 667.2 & 6 & -0.01 $\pm$ 0.09 & 195.9 & 1011.9
 & 816.0 \\ \hline
$B$ & $\alpha$ & $t_\mathrm{i}$ [d] & $t_\mathrm{f}$ [d] & D [d] & $V$ & $\alpha$ & $t_\mathrm{i}$ [d] & $t_\mathrm{f}$ [d] & D [d] \\
\end{tabular}
\end{table*}

\clearpage

\begin{longtable}{lllllll}
\caption{P\,Cygni velocities of Balmer line profiles from early time spectra of V392\,Per.}
\label{tab:PCygvels}\\
\hline Time & H$\alpha$ Slow & H$\alpha$ Fast &  H$\beta$ Slow &  H$\beta$ Fast &  H$\gamma$ Slow &  H$\gamma$ Fast \\
     / days & / km\,s$^{-1}$ & / km\,s$^{-1}$ & / km\,s$^{-1}$ & / km\,s$^{-1}$ & / km\,s$^{-1}$ & / km\,s$^{-1}$ \\
\hline
\endfirsthead

\multicolumn{7}{c}{\tablename\ \thetable{} -- continued from previous page} \\
\hline
   Time & H$\alpha$ Slow &  H$\alpha$ Fast &  H$\beta$ Slow &  H$\beta$ Fast &  H$\gamma$ Slow &  H$\gamma$ Fast \\
    / days  & / km\,s$^{-1}$ & / km\,s$^{-1}$ & / km\,s$^{-1}$ & / km\,s$^{-1}$ & / km\,s$^{-1}$ & / km\,s$^{-1}$ \\
\hline
\endhead

\hline
\multicolumn{7}{r}{Continued on next page}
\endfoot

\hline
\endlastfoot
1.91 & $-2994 \pm 23$ & $-4960 \pm 23$ & $-3209 \pm 31$ & $-5061 \pm 31$ & $-3180 \pm 35$ & $-4873 \pm 35$ \\
2.13 & $-3070 \pm 32$ & $-5053 \pm 32$ & $-3602 \pm 43$ & $-5459 \pm 43$ & $-3583 \pm 48$ & $-5035 \pm 48$ \\
2.88 & $-3748 \pm 23$ & $-5485 \pm 23$ & $-3796 \pm 31$ & $-5153 \pm 31$ & $-3802 \pm 35$ & $-5046 \pm 35$ \\
2.90  & $-3840 \pm 23$ & $-5600 \pm 23$ & $-3641 \pm 31$ & $-5153 \pm 31$ & $-3664 \pm 35$ & $-5046 \pm 35$ \\
3.85 & \ldots              & $-4068 \pm 23$ & \ldots              & $-5153 \pm 31$ & \ldots              & $-5046 \pm 35$ \\
3.87 & \ldots              & $-4092 \pm 26$ & \ldots              & $-3861 \pm 36$ & \ldots              & $-3716 \pm 40$ \\
4.16 & \ldots              & $-3829 \pm 40$ & \ldots              & $-3781 \pm 54$ & \ldots              & $-3249 \pm 61$ \\
4.80  & \ldots              & \ldots              & \ldots              & $-3905 \pm 11$ & \ldots              & \ldots              \\
\end{longtable}

\clearpage

\begin{longtable}{lllll}
\caption{Dereddened line fluxes for H$\alpha$, H$\beta$, [\ion{O}{iii}] 5007\,\AA\ and [\ion{O}{iii}] 4959\,\AA.}
\label{tab:Fluxes_all}\\
\hline
Time & H$\alpha$ Flux  & H$\beta$ Flux  & [\ion{O}{iii}] 5007\,\AA\ Flux & [\ion{O}{iii}] 4959\,\AA\ Flux  \\
/ days  & / erg\,cm$^{-2}$\,s$^{-1}$ & / erg\,cm$^{-2}$\,s$^{-1}$ & / erg\,cm$^{-2}$\,s$^{-1}$ & / erg\,cm$^{-2}$\,s$^{-1}$\\
\hline
\endfirsthead

\multicolumn{5}{c}{\tablename\ \thetable{} -- continued from previous page} \\
\hline
Time & H$\alpha$ Flux  & H$\beta$ Flux  & [\ion{O}{iii}] 5007\,\AA\ Flux & [\ion{O}{iii}] 4959\,\AA\ Flux  \\
/ days  & / erg\,cm$^{-2}$\,s$^{-1}$ & / erg\,cm$^{-2}$\,s$^{-1}$ & / erg\,cm$^{-2}$\,s$^{-1}$ & / erg\,cm$^{-2}$\,s$^{-1}$\\
\hline
\endhead

\hline
\multicolumn{5}{r}{Continued on next page}
\endfoot

\hline
\endlastfoot
4.9 & $\left( 3.219 \pm 0.060 \right) \times 10^{-7} $ & $ \left( 1.815 \pm 0.018 \right) \times 10^{-7} $ & \ldots & \ldots \\
5.9 & $ \left( 2.514 \pm 0.042 \right) \times 10^{-7} $ & $ \left( 1.119 \pm 0.018 \right) \times 10^{-7} $ & \ldots & \ldots \\
6.9 & $ \left( 2.739 \pm 0.031 \right) \times 10^{-7} $ & $ \left( 1.026 \pm 0.014 \right) \times 10^{-7} $ & \ldots & \ldots \\
76.2 & $ \left( 6.852 \pm 0.133 \right) \times 10^{-10} $ & \ldots & \ldots & \ldots \\
77.2 & $ \left( 6.811 \pm 0.118 \right) \times 10^{-10} $ & $ \left( 9.516 \pm 0.867 \right) \times 10^{-11} $ & $ \left( 9.448 \pm 0.257 \right) \times 10^{-10} $ & $ \left( 2.724 \pm 0.171 \right) \times 10^{-10} $ \\
82.2 & $ \left( 5.179 \pm 0.105 \right) \times 10^{-10} $ & $ \left( 1.426 \pm 0.061 \right) \times 10^{-10} $ & $ \left( 1.359 \pm 0.022 \right) \times 10^{-9} $ & $ \left( 4.131 \pm 0.126 \right) \times 10^{-10} $ \\
84.2 & $ \left( 4.645 \pm 0.122 \right) \times 10^{-10} $ & $ \left( 1.349 \pm 0.095 \right) \times 10^{-10} $ & $ \left( 1.336 \pm 0.034 \right) \times 10^{-9} $ & $ \left( 4.053 \pm 0.185 \right) \times 10^{-10} $ \\
87.2 & $ \left( 4.175 \pm 0.093 \right) \times 10^{-10} $ & $ \left( 1.512 \pm 0.089 \right) \times 10^{-10} $ & $ \left( 1.968 \pm 0.026 \right) \times 10^{-9} $ & $ \left( 6.061 \pm 0.197 \right) \times 10^{-10} $ \\
\end{longtable}

\clearpage

\begin{longtable}{lllll}
\caption{Dereddened line fluxes for [\ion{O}{iii}] 4363\,\AA, \ion{He}{i} 6678\,\AA, \ion{He}{i} 7065\,\AA\ and \ion{He}{ii} 4686\,\AA.}
\label{tab:Fluxes_all2}\\
\hline
Time & [\ion{O}{iii}] 4363\,\AA\ flux & \ion{He}{i} 6678\,\AA\ flux & \ion{He}{i} 7065\,\AA\ flux & \ion{He}{ii} 4686\,\AA\ flux  \\
/ days  & / erg\,cm$^{-2}$\,s$^{-1}$ & / erg\,cm$^{-2}$\,s$^{-1}$ & / erg\,cm$^{-2}$\,s$^{-1}$ & / erg\,cm$^{-2}$\,s$^{-1}$\\
\hline
\endfirsthead

\multicolumn{5}{c}{\tablename\ \thetable{} -- continued from previous page} \\
\hline
Time & [\ion{O}{iii}] 4363\,\AA\ flux & \ion{He}{i} 6678\,\AA\ flux & \ion{He}{i} 7065\,\AA\ flux & \ion{He}{ii} 4686\,\AA\ flux  \\
/ days  & / erg\,cm$^{-2}$\,s$^{-1}$ & / erg\,cm$^{-2}$\,s$^{-1}$ & / erg\,cm$^{-2}$\,s$^{-1}$ & / erg\,cm$^{-2}$\,s$^{-1}$\\

\hline
\endhead

\hline
\multicolumn{5}{r}{Continued on next page}
\endfoot

\hline
\endlastfoot
4.9 & \ldots & $ \left( 5.303 \pm 0.011 \right) \times 10^{-9} $ & $ \left( 3.735 \pm 0.038 \right) \times 10^{-8} $ & \ldots \\
5.9 & \ldots & $ \left( 2.763 \pm 0.003 \right) \times 10^{-9} $ & $ \left( 2.754 \pm 0.032 \right) \times 10^{-8} $ & \ldots \\
6.9 & \ldots & $ \left( 2.696 \pm 0.002 \right) \times 10^{-9} $ & $ \left( 2.107 \pm 0.021 \right) \times 10^{-8} $ & \ldots \\
76.2 & \ldots & $ \left( 5.266 \pm 0.051 \right) \times 10^{-12} $ & $ \left( 1.989 \pm 0.061 \right) \times 10^{-10} $ & \ldots \\
77.2 & $ \left( 1.757 \pm 0.327 \right) \times 10^{-10} $ & $ \left( 4.664 \pm 0.012 \right) \times 10^{-12} $ & $ \left( 1.694 \pm 0.032 \right) \times 10^{-10} $ & \ldots \\
82.2 & $ \left( 3.897 \pm 0.216 \right) \times 10^{-10} $ & $ \left( 2.741 \pm 0.053 \right) \times 10^{-12} $ & $ \left( 8.373 \pm 0.364 \right) \times 10^{-11} $ & $ \left( 1.089 \pm 0.102 \right) \times 10^{-10} $ \\
84.2 & $ \left( 4.595 \pm 0.289 \right) \times 10^{-10} $ & \ldots & $ \left( 6.934 \pm 0.383 \right) \times 10^{-11} $ & $ \left( 9.858 \pm 1.208 \right) \times 10^{-11} $ \\
87.2 & $ \left( 4.752 \pm 0.256 \right) \times 10^{-10} $ & $ \left( 2.436 \pm 0.010 \right) \times 10^{-12} $ & $ \left( 1.164 \pm 0.028 \right) \times 10^{-10} $ & \ldots \\
\end{longtable}

\bsp
\label{lastpage}
\end{document}